\documentclass[article]{elsarticle}

\usepackage[utf8]{inputenc}
\usepackage{amssymb}
\usepackage{amsfonts}
\usepackage{graphicx}
\usepackage{multirow}
\usepackage{bigstrut}
\usepackage{epstopdf}
\usepackage[english]{babel}
\usepackage{combelow}
\usepackage{amsmath}
\usepackage{amsthm}
\usepackage{algorithm}
\usepackage{algorithmicx}
\usepackage[noend]{algpseudocode}
\usepackage{xcolor}
\usepackage{bm}
\usepackage{siunitx,makecell,booktabs}
\usepackage{adjustbox}     
\usepackage[toc,page]{appendix} 

\usepackage{tikzpagenodes}
\usepackage{tikz}
\usetikzlibrary{shapes}
\usetikzlibrary{arrows}
\usetikzlibrary{positioning}
\usetikzlibrary{matrix}
\usetikzlibrary{patterns}
\usetikzlibrary{decorations.pathreplacing,decorations.pathmorphing,decorations.markings}
\usepackage{tikz-qtree}
\usetikzlibrary{trees,calc,arrows.meta,positioning,bending}
\usepackage{mathtools}
\usepackage{todonotes}
\usepackage{sidecap}
\usepackage{fullpage}
\usepackage{hyperref}
\usepackage{upgreek}
\usepackage[normalem]{ulem}
\usepackage[switch]{lineno} 

\newtheorem{remark}{Remark}

\graphicspath{{figures/}}
\DeclareGraphicsExtensions{.pdf,.eps,.png,.jpg,.jpeg}

\hypersetup{
    pdftitle={Fast prediction of plasma instabilities with sparse-grid-accelerated optimized dynamic mode decomposition},
    pdfauthor={Kevin Gill, Ionut-Gabriel Farcas, Silke Glas, Benjamin Faber},
    pdfkeywords={data-driven reduced-order modeling, higher-dimensional parameter spaces, higher-dimensional parametric reduced-order modeling, parametric reduced-order modeling, sparse grids, Leja points, optimized dynamic mode decomposition, plasma micro-instability simulations}
}
\usepackage{fullpage}
\usepackage{hyperref}

\definecolor{carminepink}{rgb}{0.92, 0.3, 0.26}
\definecolor{cobalt}{rgb}{0.0, 0.28, 0.67}
\definecolor{cerulean}{rgb}{0.0, 0.48, 0.65}

\newcount\Comments
\newcommand{\kibitz}[2]{\ifnum\Comments=1\textcolor{#1}{#2}\fi}

\renewcommand{\hat}[1]{\widehat{#1}}
\renewcommand{\tilde}[1]{\widetilde{#1}}

\journal{Journal of Computational Physics}

\begin{document}

\begin{frontmatter}

\title{Fast prediction of plasma instabilities with sparse-grid-accelerated optimized dynamic mode decomposition}

\author[inst1]{Kevin Gill\corref{cor1}}
\ead{kevin.gill@wisc.edu}
\cortext[cor1]{Corresponding author.}

\author[inst2]{Ionu\cb{t}-Gabriel Farca\cb{s}}
\author[inst3]{Silke Glas}
\author[inst1]{Benjamin J.~Faber}

\affiliation[inst1]{%
  organization={Department of Nuclear Engineering and Engineering Physics, University of Wisconsin--Madison},
  city={Madison},
  state={WI},
  postcode={53706},
  country={USA}
}

\affiliation[inst2]{%
  organization={Department of Mathematics and Division of Computational Modeling and Data Analytics, Academy of Data Science, Virginia Tech},
  city={Blacksburg},
  state={VA},
  postcode={24061},
  country={USA}
}

\affiliation[inst3]{%
  organization={Department of Applied Mathematics, University of Twente, P.O. Box 217},
  postcode={7500AE Enschede},
  country={The Netherlands}
}

\begin{abstract}
Parametric data-driven reduced-order models (ROMs) that embed dependencies in a large number of input parameters are crucial for enabling many-query tasks in large-scale problems. 
These tasks, including design optimization, control, and uncertainty quantification, are essential for developing digital twins in real-world applications. 
However, standard grid-based data generation methods are computationally prohibitive due to the curse of dimensionality, as their cost scales exponentially with the number of inputs.
This paper investigates efficient training of parametric data-driven ROMs using sparse grid interpolation with $(L)$-Leja points, specifically targeting scenarios with higher-dimensional input parameter spaces. 
$(L)$-Leja points are nested and exhibit slow growth, resulting in sparse grids with low cardinality in low-to-medium dimensional settings, making them ideal for large-scale, computationally expensive problems.
Focusing on gyrokinetic simulations of plasma micro-instabilities in fusion experiments as a representative real-world application, we construct parametric ROMs for the full 5D gyrokinetic distribution function via optimized dynamic mode decomposition (optDMD) and sparse grids based on $(L)$-Leja points. 
We perform detailed experiments in two scenarios: First, the Cyclone Base Case benchmark assesses optDMD ROM prediction capabilities beyond training time horizons and across variations in the binormal wave number. 
Second, for a real-world electron–temperature–gradient–driven micro-instability simulation with six input parameters, we demonstrate that a predictive parametric optDMD ROM that is up to three orders of magnitude cheaper to evaluate can be constructed using only $28$ high-fidelity gyrokinetic simulations, enabled by the use of sparse grids.
In the broader context of fusion research, these results demonstrate the potential of sparse grid-based parametric ROMs to enable otherwise intractable many-query tasks.
\end{abstract}


\end{frontmatter}

\section{Introduction} \label{sec:intro}


High-performance computing (HPC) is crucial for simulating complex real-world phenomena across scientific and engineering disciplines. 
Today's exascale supercomputers~\cite{atchleyFrontierExploringExascale2023} can handle simulations with trillions of degrees of freedom~\cite{bird2021vpic}, a scale unimaginable just decades ago. 
Historically, HPC has been used to conduct a single or a small set of forward simulations with predefined input parameters. 
Yet, addressing practically relevant tasks---such as designing and controlling optimized fusion reactors or performing numerical simulations to support the development of next-generation propulsion devices---requires moving beyond this approach. 
For tasks like design, control, sensitivity analysis, uncertainty quantification, and solving inverse problems, the cost of performing large ensembles of high-fidelity simulations is prohibitive. 
Consequently, parametric reduced-order models (ROMs)~\cite{BGW15}, which embed the variation of key input parameters with the goal of providing computationally cheap approximations, are as relevant as ever.

This paper focuses on non-intrusive parametric data-driven ROMs with a higher-dimensional parameter space trained from simulation data.
This approach is motivated by the complexity of large-scale simulation codes, making data-driven methods a convenient way to construct ROMs without needing access to potentially complex or proprietary codebases.
Given a set of training parameter instances, parametric ROMs are built using the corresponding high-fidelity simulation data.
Once trained, they provide rapid online predictions for arbitrary parameter instances, typically within the same domain as the training parameters. 
While significant effort has gone into developing new methods for constructing such data-driven parametric ROMs, less emphasis has been placed on efficiently handling high-dimensional parameter spaces.
This, in turn, is crucial in practice since the cost of a single high-fidelity simulation can be significant.
In this paper, we investigate a sparse grid approach for efficiently constructing parametric ROMs suitable for higher-dimensional parameter spaces and computationally expensive simulation codes. 
We demonstrate the effectiveness of this approach in a real-world application: the simulation of plasma micro-instabilities in fusion experiments, an area that stands to benefit greatly from parametric ROMs.
We note, however, that the ideas proposed in this work are broadly applicable and easily transferable to other application scenarios.

Achieving commercially viable fusion energy hinges on our ability to understand, predict, and control turbulent transport in magnetic confinement devices~\cite{Yoshida_2025}. 
In this regard, HPC plays a key role through detailed simulations, enabling a shift from simpler qualitative descriptions to detailed and quantitative predictions, enabling progress towards full-device~\cite{germaschewskiExascaleWholedeviceModeling2021} and complex multiscale turbulent transport simulations~\cite{belliSpectralTransitionMultiscale2022}.
However, these advanced simulations consume significant resources (e.g., millions of compute hours), limiting the number of cases that can be studied explicitly. 
This critically underscores the need for innovative parametric reduced-order modeling approaches to enable these otherwise infeasible tasks.

Over the past decades, parametric ROMs have been the subject of extensive research. 
This body of work encompasses intrusive methods~\cite{AF11, Bi11, BGHU24, BWG08b, BWG08a, FB21, grepl2007efficient, MR02, PZB13, HKP13} and non-intrusive approaches~\cite{ADR23, Ch24, Hu23, MKW23, Ra20, Re20,Sc10, Du17} developed for both linear and nonlinear systems.
Non-intrusive approaches also include more recent extensions based on machine learning~\cite{Fu23, Lee2024, RMR20}.
A comprehensive review is available in Ref.~\cite{BGW15}. 
Generally, these methods necessitate acquiring a training dataset, typically from high-fidelity simulations, for a given set of input parameter instances. 
However, when the number of parameters is large, standard grid-based methods for selecting this set of parameter instances become prohibitively expensive due to the curse of dimensionality.

Sparse grids~\cite{BG04} offer a promising approach for delaying the curse of dimensionality in higher-dimensional parameter spaces. 
Essentially, sparse grids construct a multi-dimensional grid by intelligently combining lower-dimensional full tensor product grids, prioritizing grid point combinations that contribute most significantly to overall accuracy of the target approximation.
Sparse grids were also employed in model reduction. 
For instance, Ref.~\cite{PZB13} utilized sparse grids, specifically those constructed using standard dyadic grids, with the reduced basis method~\cite{MR02} for constructing reduced basis spaces. 
Ref.~\cite{PZB13} also explored directly interpolating the output of interest on sparse grids to circumvent the reduced basis method entirely. 
In a similar vein, Ref.~\cite{Bu17} used dyadic sparse grids for parametric ROMs of large symmetric linear time-invariant dynamical systems. 
A non-intrusive reduction method for porous media multiphase flows using Smolyak sparse grids was developed in Ref.~\cite{Du17}. 
Ref.~\cite{Za19} employed stochastic collocation based on dimension-adaptive sparse grids~\cite{GG03} and projection-based ROMs for optimization problems constrained by partial differential equations (PDEs) with uncertain coefficients.
In contrast to existing approaches--which have primarily focused on intrusive reduced modeling, bypassing the ROM construction, or were limited to linear time-invariant systems or specialized applications--we propose integrating sparse grid interpolation based on $(L)$-Leja points~\cite{JWZ18,NJ14} into the parametric data-driven ROM construction.
This idea is broadly applicable to both linear and nonlinear data-driven parametric reduced modeling.
$(L)$-Leja points can be constructed over arbitrary domains characterized by arbitrary density functions, are arbitrarily granular and form a nested interpolatory sequence (i.e., only one extra point is needed to increase the interpolation degree by one). 
This means that the cardinality of sparse $(L)$-Leja grids grows slowly with the dimension, which is desirable for computationally expensive problems. 
Furthermore, these points were shown to have excellent interpolation accuracy~\cite{NJ14}.
Here, we will use sparse $(L)$-Leja grids to generate training data and for interpolation to compute parametric ROM predictions for arbitrary input parameter instances. 

We will demonstrate the potential of our parametric data-driven reduced modeling approach with $(L)$-Leja sparse grids in plasma micro-instability simulations~\cite{Garbet_2010}.
We model plasma micro-instabilities via the gyrokinetic system of equations~\cite{Brizard2007}, which represents perhaps the most widely used approach in practice.
Gyrokinetics effectively reduces the plasma phase space from 6D (three positions, three velocities) to 5D (three positions, two velocities) and allows for larger numerical step sizes by averaging over the fast gyro-motion of plasma particles.
Various simulation approaches exist to solve the gyrokinetic equations~\cite{barnes_2024_14022029, Yang2007,  Hakim2020, Jenko_GENE, KLEIBER2024, Ku2018,LANTI2020, Lin1998,  Mandell2024, Watanabe_2006}.
Here, we will use the gyrokinetic code \textsc{Gene}~\cite{Goe11, Jenko_GENE}, an Eulerian (i.e., grid-based) code that implements one of the most comprehensive physical models for studying micro-instabilities and micro-turbulence. 
With recent advancements in data-driven modeling and machine learning, there has been a growing interest in surrogate modeling for plasma physics~\cite{Artigues_2025, farajiDynamicModeDecomposition2023, farajiDynamicModeDecomposition2023II, Fr23,Gopakumar_2024, kaptanogluCharacterizingMagnetizedPlasmas2020, vPl20, Ta18}, including gyrokinetics~\cite{Du25, FMJ24, galletti20255dneuralsurrogatesnonlinear, Hatch_2024, Hatch_2022,Ro22}.
Moreover, sparse grids have been leveraged to efficiently construct surrogate models for the mapping between key input parameters and scalar outputs~\cite{FMJ22,FDJ21, FMJ24, Ko22}.

Our goal is to go beyond standard approaches and construct parametric data-driven ROMs to quickly and reliably predict the \emph{full} phase space structure of dominant micro-instability modes in scenarios with a higher-dimensional parameter space.
This is achieved by performing linear (in phase space variables) simulations, which are essential for identifying the underlying micro-instabilities.
To this aim, we will construct parametric data-driven ROMs based on optimized Dynamic Mode Decomposition (optDMD)~\cite{askham2018variable, CTR12} using training data computed with \textsc{Gene}.
Standard DMD~\cite{Ku16, Sc10, Tu14} is a data-driven dimensionality reduction technique, developed by Peter Schmid~\cite{Sc10}, that extracts meaningful spatiotemporal patterns from data. 
It identifies a set of dynamic modes, each associated with a specific oscillation frequency and a growth or decay rate. 
DMD seeks to find a best-fit linear operator that maps a data snapshot at one time instant to the next, making it an ideal candidate for modeling linear systems (and beyond, due to its connection to Koopman theory~\cite{RMBSH09}). 
DMD is often used for analysis, i.e., for understanding the underlying dynamic processes of a system from observed data, as demonstrated in a recent study on mode transition for electromagnetic drift wave instabilities in linear gyrokinetic simulations~\cite{Du25}.
However, standard DMD faces challenges such as sensitivity to noise and difficulty in selecting the optimal number of modes for reconstruction. 
In contrast, optDMD addresses these limitations by recasting the DMD problem as a nonlinear optimization problem to find the optimal set of DMD modes and their corresponding dynamics.
Here, we will go beyond conventional analysis as well as surrogate models that map input parameters to scalar outputs and construct parametric data-driven ROMs that integrate optDMD and sparse grid interpolation based on $(L)$-Leja points for predicting the full phase space structure of micro-instability modes across a wide range of input parameters in fusion plasma scenarios.
In a real-world scenario involving electron temperature gradient-driven micro-instability modes at the edge of fusion experiments~\cite{Dorland, Jenko_GENE}, simulated with \textsc{Gene} in a six-dimensional parameter space, we show that a predictive parametric sparse grid optDMD ROM can be trained using only $28$ $(L)$-Leja points.
Our code is available at~\url{https://github.com/kevinsinghgill/parametric_ROMs}.
To our knowledge, this represents one of the first investigations into parametric reduced modeling specifically for gyrokinetic simulations of plasma micro-instabilities. 
In the broader context of fusion research, parametric ROMs represent a tool for enabling otherwise infeasible tasks such as (design) optimization, parameter exploration, and uncertainty quantification.

The remainder of this paper is organized as follows. 
Section~\ref{sec:background} provides background on data-driven reduced modeling with parametric variations, optDMD, and sparse grid interpolation using $(L)$-Leja points. Section~\ref{sec:background_gyrokinetics} summarizes gyrokinetic modeling for plasma micro-instability simulations and the gyrokinetic code \textsc{Gene}. 
Our approach for constructing parametric data-driven ROMs for higher-dimensional input spaces with optDMD and $(L)$-Leja sparse grid interpolation is presented in Section~\ref{sec:proposed_approach}. 
Section~\ref{sec:results} presents our findings across two plasma micro-instability simulation scenarios: the Cyclone Base Case benchmark~\cite{Di00}, where we investigate optDMD ROMs for forecasting beyond a training time horizon and for variations in a key plasma parameter, and a real-world  electron temperature gradient mode simulation with variations in six parameters. 
Section~\ref{sec:conclusion} concludes the paper.

\section{Preliminaries}
\label{sec:background}

Section~\ref{subsec:background_datadrivenrom} summarizes the general setup for data-driven parametric reduced modeling. 
We then provide overviews of optDMD (Sec.~\ref{subsec:background_optdmd}) and sparse grid interpolation with $(L)$-Leja points (Sec.~\ref{subsec:background_SG_interp}), which form the building blocks for our parametric data-driven reduced modeling approach.

\subsection{Setup for data-driven reduced-order modeling with parametric variations}
\label{subsec:background_datadrivenrom}

This paper focuses on learning parametric data-driven ROMs that embed the variation in $d \in \mathbb{N}$ input parameters $\pmb{\uptheta} \in \mathcal{D}$ in a domain $\mathcal{D} \subseteq \mathbb{R}^d$.
We specifically target scenarios where the number of parameters, $d$, is sufficiently large to warrant efficient approaches for generating training data to construct ROMs.
Let $[t_{i}, t_{f}]$ denote a time interval of interest, with $t_{i}$ being the initial time and $t_{f}$ being the final time.
We define the complex-valued state vector $\mathbf{g}(t; \pmb{\uptheta}) = [g_1(t; \pmb{\uptheta}), g_2(t; \pmb{\uptheta}), \ldots, g_n(t; \pmb{\uptheta})]^\top \in \mathbb{C}^n$ of dimension $n \in \mathbb{N}$. 
The dynamics are generically modeled as a system of ordinary differential equations (ODEs), i.e., 
\begin{equation} \label{eq:FOM_general}
    \frac{\mathrm{d}\mathbf{g}}{\mathrm{d} t}(t; \pmb{\uptheta}) = \mathbf{f}(\mathbf{g}(t;\pmb{\uptheta}), t; \pmb{\uptheta}), \quad \mathbf{g}(t_{i}; \pmb{\uptheta})=\mathbf{g}_{\mathrm{init}}( \pmb{\uptheta}),
\end{equation}
where $\mathbf{g}_{\mathrm{init}}( \pmb{\uptheta})$ is a specified initial condition and $\mathbf{f} : \mathbb{C}^{n} \times [t_{i}, t_{f}] \times \mathcal{D} \rightarrow \mathbb{C}^{n}$ is a (possibly nonlinear) function that defines the time evolution of the system state. 
In general, this ODE system stems from the spatial discretization of the governing PDEs, with dimension $n$ that scales with the size of the spatial discretization.
In plasma micro-instability and micro-turbulence simulations, $n$ is typically of order $10^5$--$10^9$. 

Let $\pmb{\uptheta}_1, \pmb{\uptheta}_2, \ldots, \pmb{\uptheta}_{n_p} \in \mathcal{D}$ be $n_p \in \mathbb{N}$ parameter instances and let $\mathbf{g}_k(\pmb{\uptheta}_j)$ denote the state solution at time $t_k$ for parameter instance $\pmb{\uptheta}_j$, where we use the notation $\mathbf{g}_k(\pmb{\uptheta}_j) \equiv \mathbf{g}(t_k; \pmb{\uptheta}_j)$.
We collect $n_t \in \mathbb{N}$ uniformly spaced snapshots over the time domain of interest $[t_i, t_f]$ for each parameter instance for each $\pmb{\uptheta}_1, \pmb{\uptheta}_2, \ldots, \pmb{\uptheta}_{n_p}$ by solving the high-fidelity model \eqref{eq:FOM_linear_gyrokinetics} and recording the corresponding solutions at the $n_t$ time instants.
Note that we assume that we collect the \emph{same} number of snapshots $n_t$ for each training parameter instance.
For a training parameter instance $\pmb{\uptheta}_j$, we define the trajectory 
\begin{align*}
\mathbf{G}_j := \left[ \mathbf{g}_1(\pmb{\uptheta}_j), \mathbf{g}_2(\pmb{\uptheta}_j), \ldots, \mathbf{g}_{n_t}(\pmb{\uptheta}_j) \right] \in \mathbb{C}^{n \times n_t}.
\end{align*}
In practice, the matrices $\mathbf{G}_j$ are ``tall-and-skinny", i.e., $n \gg n_t$.
Given $\mathbf{G}_1, \mathbf{G}_2, \ldots, \mathbf{G}_{n_p}$, our goal is to learn parametric data-driven ROMs that provide rapid and reliable predictions for an arbitrary $\bm{\uptheta} \in \mathcal{D}$.

Several methods exist for constructing parametric ROMs that embed parametric dependence; a comprehensive survey can be found in Ref.~\cite{BGW15}. 
These methods generally differ in how they build the reduced basis and the reduced model operators.
When it comes to constructing the reduced basis, approaches typically fall into two categories: those that use local bases (built for each training parameter instance $\pmb{\uptheta}_1, \pmb{\uptheta}_2, \ldots, \pmb{\uptheta}_{n_p}$) and those that employ a global reduced basis (constructed using data across all training parameter instances). 
Similarly, the reduced model operators can be built for each instance of the training parameters, or they can be constructed as a single, global set applicable to all parameters.
The latter is particularly well-suited for models with an affine dependence on $\pmb{\uptheta}$.
For example, for a linear time-invariant model described by $\frac{\mathrm{d}\mathbf{g}}{\mathrm{d} t}(t; \pmb{\uptheta}) = \mathbf{A}(\mathbf{g}; \pmb{\uptheta})$, an affine structure means that $\mathbf{A}(\mathbf{g}; \pmb{\uptheta}) = \sum_{k=1}^{q_A} \nu_k(\pmb{\uptheta}) \mathbf{A}_k(\mathbf{g})$, where $q_A$ is the number of affine terms and $\nu_k(\pmb{\uptheta})$ are scalar-valued functions such that $\{\nu_k(\pmb{\uptheta})\}_{k=1}^{q_A}$ is linearly independent.
When bases or reduced operators are individually constructed for each training parameter instance, interpolation is commonly used to derive the corresponding reduced basis or operator for any arbitrary $\bm{\uptheta} \in \mathcal{D}$.
In this work, we compute a global reduced basis using the data for all training parameter instances and use interpolation to compute predictions for an arbitrary $\pmb{\uptheta} \in \mathcal{D}$.
More details are provided in Sec.~\ref{sec:proposed_approach}.

\subsection{Learning data-driven reduced models with optimized dynamic mode decomposition}
\label{subsec:background_optdmd}
The target application in the present paper is the simulation of plasma micro-instabilities.
Since these simulations are linear in the phase-space variables, linear ROMs are a natural choice.
Accordingly, we employ optDMD to construct the parametric ROMs.
This section provides a brief overview of standard DMD, followed by a summary of optDMD (for further details, see Refs.~\cite{askham2018variable,CTR12,Ku16,Sc10,Tu14}). 
Subsequently, Sec.~\ref{sec:proposed_approach} will detail how optDMD is specifically employed to construct parametric ROMs.

Standard DMD is a data-driven method used to decompose complex system dynamics into a set of spatiotemporal modes, each associated with a fixed oscillation frequency and decay or growth rate. 
It is particularly effective for analyzing time-resolved data from simulations or experiments.
Assume the parameter of interest is fixed to a value $\pmb{\uptheta} = \pmb{\uptheta}_f$ and let $\mathbf{G} \equiv \mathbf{G}_f = \begin{bmatrix} \mathbf{g}_1(\pmb{\uptheta}_f), \mathbf{g}_2(\pmb{\uptheta}_f), \ldots, \mathbf{g}_{n_t}(\pmb{\uptheta}_f) \end{bmatrix} \in \mathbb{C}^{n \times n_t}$ denote the corresponding snapshot matrix.
Standard DMD seeks to find a best-fit linear operator $\mathbf{A} \equiv \mathbf{A}(\mathbf{g}; \pmb{\uptheta}_f) \in \mathbb{C}^{n \times n}$ such that $\mathbf{g}_{k+1}(\pmb{\uptheta}_f) \approx \mathbf{A}\mathbf{g}_k(\pmb{\uptheta}_f)$. 
This operator $\mathbf{A}$ maps the system state from one time step to the next.
Arranging the snapshot data into the following two shifted matrices $\mathbf{G}^{\ell}, \mathbf{G}^{\ell +1} \in \mathbb{C}^{n \times (n_t-1)}$\,,
\begin{align*}
    \mathbf{G}^{\ell} = \begin{bmatrix} \mathbf{g}_{1}(\pmb{\uptheta}_f) & \mathbf{g}_{2}(\pmb{\uptheta}_f) & \dots & \mathbf{g}_{n_t-1}(\pmb{\uptheta}_f) \end{bmatrix}, \quad
    \mathbf{G}^{\ell +1} = \begin{bmatrix} \mathbf{g}_{2}(\pmb{\uptheta}_f) & \mathbf{g}_{3}(\pmb{\uptheta}_f) & \dots & \mathbf{g}_{n_t}(\pmb{\uptheta}_f) \end{bmatrix},
\end{align*}
the linear operator $\mathbf{A}$ can be approximated by minimizing the Frobenius norm of the residual:
$$ \min_{\mathbf{A}} \|\mathbf{G}^{\ell +1} - \mathbf{A}\mathbf{G}^{\ell}\|_F .$$
The solution to this least-squares problem is $\mathbf{A} = \mathbf{G}^{\ell +1} \left(\mathbf{G}^{\ell}\right)^{\dagger}$, where $\left(\mathbf{G}^{\ell}\right)^{\dagger}$ is the Moore-Penrose pseudoinverse of $\mathbf{G}^{\ell}$.
In large-scale applications, a low-rank approximation of $\mathbf{A}$ is often computed using the singular value decomposition (SVD) of $\mathbf{G}^{\ell} = \mathbf{U}^{\ell} \mathbf{\Sigma}^{\ell} \left(\mathbf{V}^{\ell}\right)^*$. 
The DMD modes are then typically computed depending on the eigenvectors of a rank-$r$ projected matrix $\widehat{\mathbf{A}} = \left( \mathbf{U}^{\ell}_r \right)^* \mathbf{A} \mathbf{U}^{\ell}_r = \left(\mathbf{U}^{\ell}_r\right)^* \mathbf{G}^{\ell +1} \mathbf{V}^{\ell}_r \left(\mathbf{\Sigma}^{\ell}_r\right)^{-1}$. 
To simplify the notation, we omit the dependence on $\pmb{\uptheta}_f$ in the remainder of this section.

Let $\widehat{\pmb{\uplambda}} \in \mathbb{C}^{r}$ denote the eigenvalues of $\widehat{\mathbf{A}}$ and $\widehat{\mathbf{W}} \in \mathbb{C}^{r \times r}$ the corresponding eigenvectors.
The DMD modes $\widehat{\mathbf{\Phi}} \in \mathbb{C}^{n \times r}$ can be computed as 
$\widehat{\mathbf{\Phi}} = \mathbf{U}^{\ell}_r \widehat{\mathbf{W}}$.
The system dynamics (i.e., the distribution function for a gyrokinetic micro-instability simulation) at a given time $t$ can be expressed as:
\begin{equation} \label{eq:DMD_pred_time}
\mathbf{g}(t) = \widehat{\bm{\Phi}}\exp{(\mathrm{diag}(\widehat{\pmb{\upalpha}})t)} \widehat{\mathbf{b}},
\end{equation}
where $\widehat{\pmb{\upalpha}}$ denote the continuous time eigenvalues with entries $\upalpha_i = \ln(\widehat{\uplambda}_i)/\Delta t$ and 
\begin{equation*}
\widehat{\mathbf{b}} = \widehat{\mathbf{\Phi}}^\dagger \mathbf{g}_1
\end{equation*}
are the mode amplitudes determined from initial conditions.

Standard DMD approximates a linear operator via least-squares, which does not directly optimize the accuracy of the reconstructed dynamics.  
The optimized DMD (optDMD)~\cite{askham2018variable} addresses this limitation by reformulating the problem as a nonlinear optimization that directly minimizes the reconstruction error of the entire trajectory.
To avoid having to solve the nonlinear optimization in the full, high-dimensional space, we adopt the variable-projection-based approximate optDMD procedure from Ref.~\cite{askham2018variable}, which solves
\begin{equation*} \label{eq:optdmd_optimization}
\min_{\widehat{\pmb{\alpha}},  \widehat{\mathbf{B}}} \left\| \hat{\mathbf{G}}^\top - \mathbf{M}(\widehat{\pmb{\alpha}}) \widehat{\mathbf{B}} \right\|_F \,,
\end{equation*}
where $\hat{\mathbf{G}} = \mathbf{U}_r^* \mathbf{G} \in \mathbb{C}^{r \times n_t}$ is the reduced representation of the snapshot matrix $\mathbf{G}$, with $\mathbf{U}_r$ containing its first $r$ singular vectors.  
The matrix $\mathbf{M}(\widehat{\pmb{\upalpha}}) \in \mathbb{C}^{n_t \times r}$ has entries $(i,j)$ given by $\exp(\alpha_i t_j)$, where $t_j = j \Delta t$, and $\widehat{\mathbf{B}} \in \mathbb{C}^{r \times r}$.  
The DMD modes and initial amplitudes are then computed as
\begin{equation*}
\widehat{\bm{\Phi}}_{:,j} = \frac{1}{\widehat{b}_j}\,\mathbf{U}_r \widehat{\mathbf{B}}^{\top}_{:,j}, \qquad
\widehat{b}_j = \|\widehat{\mathbf{B}}^{\top}_{:,j}\|_2,\quad j=1,\dots,r,
\end{equation*}
while $\widehat{\pmb{\upalpha}}$ correspond to the DMD eigenvalues.
Using the computed modes, eigenvalues, and initial amplitudes, we can evaluate the system dynamics at time $t$ via Eq.~\ref{eq:DMD_pred_time}.

In practice, the non-convex reduced optDMD optimization problem is solved using a Levenberg-Marquardt (LM) trust-region method~\cite{askham2018variable}.
As remarked in~\cite{askham2018variable}, it is not guaranteed that this will converge to the global minimizer.
Indeed, given an initial guess $\widehat{\pmb{\upalpha}}^{(0)}$, the optDMD algorithm fundamentally functions as a postprocessor that computes a nearby local minimizer to improve upon $\widehat{\pmb{\upalpha}}^{(0)}$.
To guide the search towards a physically relevant solution, we follow the recommendation in Ref.~\cite{askham2018variable} and initialize the optDMD procedure with $\widehat{\pmb{\upalpha}}^{(0)} = \ln(\widehat{\pmb{\uplambda}}^{(0)})/\Delta t$, where $\widehat{\pmb{\uplambda}}^{(0)}$ are the discrete eigenvalues computed via standard DMD.   
In this work, we employ a single start (no multi-start or bagging), using \texttt{pyDMD}'s~\cite{Ichinaga2024} default variable-projection settings, while monitoring the relative Frobenius reconstruction error and the solver's convergence. 
If a run fails to reduce the objective or triggers a non-convergence warning, we lower the truncation rank as a fallback. 
Crucially, across all experiments reported in the present paper, the single-start LM initialized from standard DMD consistently converged and produced the debiased behavior expected of optDMD.

\subsection{Sparse grid interpolation using $(L)$-Leja points}
\label{subsec:background_SG_interp}
A key challenge in parametric reduced modeling is the efficient selection of training parameter instances when the number of parameters, $d$, is large.
Standard grid-based sampling strategies suffer from the curse of dimensionality, making them impractical for high-dimensional parameter spaces.
To address this challenge within our data-driven parametric ROM framework, we integrate sparse grids—specifically those constructed from $(L)$-Leja points, into both the training and prediction phases. 
In our approach, the $(L)$-Leja points determine the parameter instances at which ROMs are trained, and sparse-grid interpolation is then used to obtain ROM predictions at arbitrary parameter values within $\mathcal{D}$.
In particular, we interpolate the optDMD eigenvalues, modes, and initial amplitudes, as detailed in Sec.~\ref{sec:proposed_approach}.  
To give the reader the necessary context for this integration, we now summarize the sparse-grid interpolation framework based on $(L)$-Leja points.

Let $h : \mathcal{X} \subseteq \mathbb{R}^d \rightarrow \mathbb{R}$ denote a multivariate function that is at least continuous on $\mathcal{X}$.
Furthermore, let $\pi$ denote a multivariate probability density function with support $\mathcal{X}$.
We assume the components of the input $\mathbf{x} \in \mathcal{X}$ to $h$ are independent.
This implies that $\mathcal{X}$ and $\pi$ must have a product structure, i.e., $\mathcal{X} := \bigotimes_{i=1}^d \mathcal{X}_i$ and $\pi(\textbf{x}) := \prod_{i=1}^d \pi_i(x_i)$, where $\mathcal{X}_i$ is the support of the density $\pi_i$ associated with input $x_i$.

Sparse grid interpolation aims to provide a computationally efficient approach for approximating $h$ at an arbitrary $\mathbf{x} \in \mathcal{X}$.
This is achieved by forming a linear combination of judiciously selected $d$-variate products of one-dimensional approximations, formally written as follows:
\begin{equation} \label{eq:tensor_delta_finite}
\mathcal{U}^{d}_{\mathcal{L}}[h]:\mathcal{X} \rightarrow \mathbb{R}, \quad \mathcal{U}^{d}_{\mathcal{L}}[h](\mathbf{x}) = \sum_{\boldsymbol{\ell} \in \mathcal{L}} \Delta^{d}_{\boldsymbol{\ell}}[h](\mathbf{x}) = \sum_{\boldsymbol{\ell} \in \mathcal{L}} \sum_{\mathbf{z} \in \{0, 1\}^{d}} (-1)^{|\mathbf{z}|_1} \mathcal{U}^{d}_{\boldsymbol{\ell} - \mathbf{z}}[h](\mathbf{x}),
\end{equation}
where $\boldsymbol{\ell} = (\ell_1, \ell_2, \ldots \ell_{d}) \in \mathbb{N}^{d}$ denotes a \emph{multiindex}, and $\mathcal{L} \subset \mathbb{N}^d$ is a \emph{multiindex set} defined as
\begin{equation} \label{eq:std_sg_index_set}
\mathcal{L} = \{\boldsymbol{\ell} \in \mathbb{N}^{d}: |\boldsymbol{\ell}|_{1} \leq L  + d - 1 \},
\end{equation}
where $|\boldsymbol{\ell}|_{1} = \sum_{i=1}^d \ell_i$ and $L \in \mathbb{N}$ denotes the sparse grid level.
The terms $\Delta^{d}_{\boldsymbol{\ell}}[h](\mathbf{x})$ are known as hierarchical surpluses, derived from full-grid operators, $\mathcal{U}^{d}_{\boldsymbol{\ell}}$, which are in turn are computed through the tensor product of one-dimensional approximations $\mathcal{U}^{i}_{\ell_1}, \mathcal{U}^{i}_{\ell_2}, \ldots, \mathcal{U}^{i}_{\ell_d}$ with support $\mathcal{X}_i$:
\begin{equation*} 
\mathcal{U}^{d}_{\boldsymbol{\ell}}[h](\mathbf{x}) = \left(\bigotimes_{i=1}^{d} \mathcal{U}^{i}_{\ell_i}\right)[h](\mathbf{x}), \quad 
\end{equation*}

In this work, the one-dimensional approximations, $\mathcal{U}^{i}_{\ell_i}$, are interpolation approximations based on Lagrange polynomials computed using the (first form of the) barycentric formula for improved numerical stability~\cite{BT04}.
A critical aspect for obtaining efficient and accurate interpolation approximations is the choice of interpolation points or knots.
Here, we employ $(L)$-Leja points which were shown to have an excellent interpolation accuracy~\cite{NJ14}.
Moreover, these points can be constructed over arbitrary domains characterized by arbitrary probability density functions $\pi_i$.
This generality is advantageous for model reduction, as it imposes no restrictions on the parameter domains or input distributions that can be treated.
Let $N_i$ denote the number of 1D $(L)$-Leja points associated to index $\ell_i$. 
The one-dimensional sequence $\{x_1, x_2, \ldots, x_{N_i} \} \sim \pi_i$ of $(L)$-Leja points is constructed greedily as:
\begin{equation*}
\begin{split}
& x_1 = \underset{x \in \mathcal{X}_i}{\mathrm{argmax}} \ \sqrt{\pi_i(x)} \\
\ & x_k = \underset{x \in \mathcal{X}_i}{\mathrm{argmax}} \ \sqrt{\pi_i(x)}\prod_{j=1}^{k-1} \left| (x - x_j) \right|, \quad k = 2, 3, \ldots, N_i.
\end{split}
\end{equation*}
$(L)$-Leja points are nested and only one extra point is needed per level to increase the interpolation degree by one.
In this case, $N_i = \ell_i$, which we are also using in our experiments.
However, since $(L)$-Leja points are arbitrarily granular, the user has a fine-grained control over their growth beyond this choice.

\begin{figure}[htbp!]
  \centering 
  \includegraphics[width=0.7\textwidth]{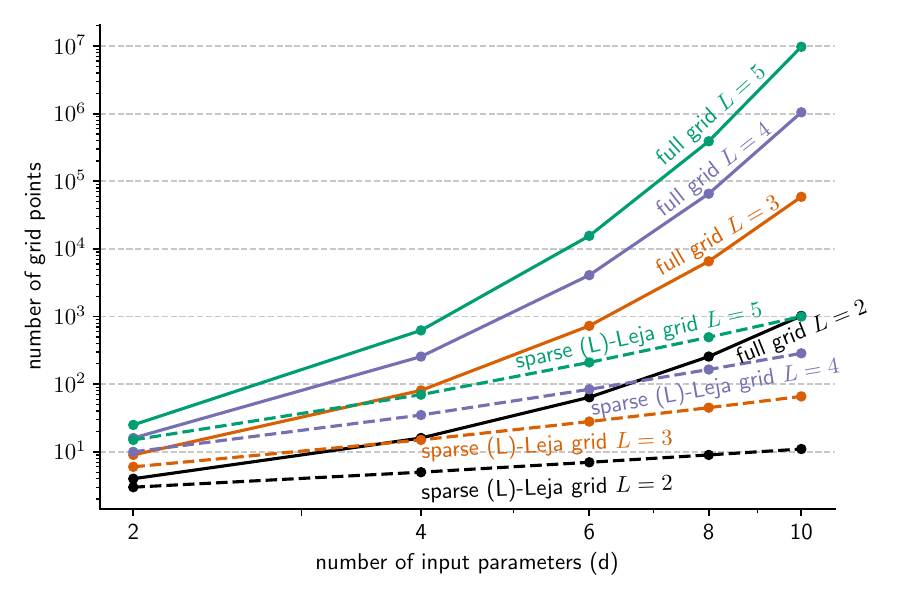}
  \caption{Cardinality of sparse $(L)$-Leja grids vs.~corresponding full tensor grid for $d\in\{2, 4, 6, 8, 10\}$ and level $L \in \{2, 3, 4, 5\}$.} 
  \label{fig:sg_vs_fg}
\end{figure}

For a given sparse grid level $L$, the multivariate interpolation approximation~\eqref{eq:tensor_delta_finite} is obtained using $N$ data pairs $\{(\mathbf{x}_j, h(\mathbf{x}_j) \}_{j=1}^N$, where $N$ denotes the total number of multivariate $(L)$-Leja points associated with the multiindex set $\mathcal{L}$ defined in Eq.~\eqref{eq:std_sg_index_set}.
$N$ is significantly smaller than the cardinality of the corresponding full-tensor grid, comprising $\prod_{i=1}^d N_i$ points.
This is shown in Fig.~\ref{fig:sg_vs_fg}, which compares the cardinality of a sparse $(L)$-Leja grid to that of the corresponding full tensor grid for $d\in\{2, 4, 6, 8, 10\}$ and $L \in \{2, 3, 4, 5\}$.
For more details about this method, we refer the reader to~\cite{FMJ22, Farcas2020, JWZ18, NJ14}.

\section{Gyrokinetic plasma micro-instability and micro-turbulence simulations}
\label{sec:background_gyrokinetics}
As a representative real-world application poised to benefit from parametric data-driven reduced modeling, we consider the simulation of plasma micro-instabilities in fusion experiments \cite{Garbet_2010}.
Section~\ref{subsec:background_gyrokinetic_modeling} summarizes gyrokinetic modeling for plasma micro-instability simulations and Sec.~\ref{subsec:background_gyrokinetics_gene} presents the gyrokinetic code \textsc{Gene} used to generate the high-fidelity training datasets for developing parametric ROMs.

\subsection{Gyrokinetic modeling for plasma micro-instability and micro-turbulence simulations}
\label{subsec:background_gyrokinetic_modeling}
Perhaps the most widely used approach for simulating plasma micro-instabilities and micro-turbulence in practice is the gyrokinetic approach~\cite{Brizard2007}, where a 5D plasma phase space distribution function $g_s(\mathbf{X},v_\|,\mu,t)$ for each plasma species $s$ is evolved in time and coupled self-consistently with external and internally generated electromagnetic fields via the coupled Vlasov-Maxwell system of PDEs.
Here, $\mathbf{X}$ is the gyrocenter position, $v_{\|}$ is the velocity component parallel to the background magnetic field $\mathbf{B}=\mathrm{B}\mathbf{b}$, and $\mu$ is the magnetic moment (a conserved quantity in the collisionless limit).
Gyrokinetics provides a good balance between the amount of physics it incorporates and its potential to be transformed into predictive, high-fidelity numerical models that can be simulated on current computing resources.
For most of plasma turbulence applications in magnetically-confined fusion plasmas, the relation $g_{1, s}/g_{0,s} \ll 1$ is satisfied, where $g_{1,s}$ denotes the \emph{fluctuating} distribution function and $g_{0,s}$ is a background (typically a static Maxwellian) distribution such that $g_s = g_{0,s} + g_{1,s}$ and $\langle g_{0,s}\rangle = 0$ for some appropriate spatio-temporal averaging operator $\langle \cdot \rangle$.
Thus many of the important properties, such as micro-instability and transport, can be examined by simulating only the evolution of $g_{1,s}$.

The dynamics of plasma species $s$ are governed by the following phase-space conservation equation:
\begin{equation}
\frac{\partial g_s}{\partial t}+\dot{\mathbf{X}}\cdot\nabla g_s+\dot{v}_{\parallel}\frac{\partial g_s}{\partial v_\|}=\sum\limits_{s'}\mathcal{C}(g_s, g_{s'}).
\label{eq0}
\end{equation}
$\mathcal{C}(g_s,g_{s'})$ denotes a collision operation describing inter- and intra-species interactions. 
The equations of motion for the gyrocenter of a particle with mass $m$ and charge $q$ read
\begin{equation*}
\dot{\mathbf{X}}=v_{\parallel}\mathbf{b}+\frac{B}{B^*_{\parallel}}\left(\mathbf{v}_{\nabla  B}+\mathbf{v}_{\kappa}+\mathbf{v}_E\right)\,, \quad
\dot{v}_{\parallel}=-\frac{\dot{\mathbf{X}}}{mv_{\parallel}}\cdot\left(\mu\nabla B +q\nabla\bar{\phi}\right) -\frac{q}{m}\dot{\bar{A}}_\parallel\,,
\end{equation*}
where $\mathbf{v}_{\nabla B}=(\mu/(m\Omega B))\,\mathbf{B}\times\nabla B$ is the grad-B drift velocity, $\mathbf{v}_{\kappa}=(v_{\parallel}^2/\Omega)\,\left(\nabla\times\mathbf{b}\right)_{\perp}$ is the curvature drift velocity, and $\mathbf{v}_E=(1/B^2)\,\mathbf{B}\times\nabla(\bar\phi-v_\|\bar{A}_\|)$ is the generalized $\mathbf{E}\times\mathbf{B}$ drift velocity. 
$\Omega=qB/m$ denotes the gyrofrequency and $B^*_{\parallel}$ is the parallel component of the effective magnetic field $\mathbf{B}^*=\mathbf{B}+\frac{B}{\Omega}v_{\parallel}\nabla\times \mathbf{b}+\nabla\times\left(\mathbf{b}\bar{A}_{\parallel}\right)$. 
In all of the above expressions, $\bar\phi$ and $\bar{A}_\parallel$ denote the gyroaveraged versions of the fluctuating electrostatic potential and the parallel component of the vector potential, which are self-consistently computed from the fluctuating distribution function. 
Assuming a static background distribution function $g_{0,s}(\mathbf{X},v_{\|},\mu)$, which allows for the decomposition $g_s(\mathbf{X},v_{\|},\mu,t)=g_{0,s}(\mathbf{X},v_{\|},\mu)+g_{1,s}(\mathbf{X},v_{\|},\mu,t)$, $\phi$ can be calculated via the Poisson equation, which, expressed at the particle position $\mathbf{x}$, reads
\begin{align*}
  \nabla^2_{\perp}\phi(\mathbf{x})=-\frac{1}{\epsilon_0}\sum_sq_s n_{1,s}(\mathbf{x})=-\frac{1}{\epsilon_0}\sum_s\frac{2\pi q_s}{m_s}\int{B^*_\parallel g_{1,s}(\mathbf{x},v_{\|},\mu)dv_\parallel d\mu}.
\end{align*}
${A}_{\parallel}$ is obtained by solving the parallel component of Amp\`ere's law for the fluctuating fields:
\begin{equation*}
-\nabla^2_{\perp}A_{\|}(\mathbf{x})=\mu_0\sum_sj_{\parallel,s}(\mathbf{x})=\mu_0\sum_s\frac{2\pi q_s}{m_s}\int{B^*_\parallel v_\parallel g_{1,s}(\mathbf{x},v_{\|},\mu)dv_\parallel d\mu}.
\end{equation*}
Expressions for connecting $g_s(\mathbf{X},v_{\|},\mu)$ and $g_s(\mathbf{x},v_{\|},\mu)$ can be found in the literature \cite{Brizard2007}.
Note that in these formulas, the time dependence has been suppressed for simplicity.

\subsection{The gyrokinetic code \textsc{Gene}}
\label{subsec:background_gyrokinetics_gene}

To perform high-fidelity gyrokinetic simulations and generate data for training parametric ROMs, we will use the plasma micro-instability and micro-turbulence simulation code \textsc{Gene}\footnote{\url{http://genecode.org}}~\cite{Goe11,Jenko_GENE}. 
\textsc{Gene} has been among the first gyrokinetic codes to leverage an Eulerian (i.e., grid-based) representation of the plasma species distributions, thus mitigating the problem of numerical noise affecting codes based on the alternative particle-in-cell approach. 
\textsc{Gene} uses a method-of-lines approach to reduce the nonlinear integro-differential gyrokinetic equations to a set of ODEs. 
In this approach, the distribution functions for all the particle species characterizing the plasma are evolved on a fixed grid in the underlying $5$D phase space.
Periodic boundary conditions are used for both perpendicular directions $x$ and $y$, which are therefore numerically implemented with spectral methods.
We denote by $n_{k_x}$ and $n_{k_y}$ the corresponding number of Fourier modes.
The total grid size is $n_{k_x}\times n_{k_y}\times n_{z}\times n_{v_\|}\times n_{\mu} \times n_s$, where $n_{z}$ denotes the number of points in the $z$ direction, and $n_{v_\|}$  $n_{\mu}$ denote the number of grid points used to discretize the velocity space.
Moreover, $n_s$ denotes the number of plasma species.
\textsc{Gene} employs a coordinate system that is aligned to the background magnetic field in configuration space to allow for an efficient representation of the turbulent fields.
This is because the turbulent fluctuations are typically field-aligned with large correlation lengths along the field lines due to parallel streaming, but are typically constrained to correlation lengths on the order of the gyroradius in the perpendicular directions. 
Furthermore, block-structured grids are adopted to optimally discretize the velocity space~\cite{Jarema_16}. 
These methods are implemented numerically through an MPI-based domain decomposition approach and achieves excellent performance over thousands of nodes \cite{Goe11}.
Recently, \textsc{Gene} has been extended to efficiently utilize GPU hardware to accelerate computationally intensive kernels, again achieving excellent performance over hundreds of nodes~\cite{germaschewskiExascaleWholedeviceModeling2021}.

\textsc{Gene} currently implements one of the most comprehensive physical models for studying micro-turbulence: nonlinear kinetic dynamics for an arbitrary number of ion species (including impurities) and electrons, electrostatic and electromagnetic fluctuation fields, interfaces with magnetohydrodynamics (MHD) codes for providing realistic magnetic equilibria \cite{Xanthopoulos2009}, as well as inter- and intra-species collision operators \cite{CRANDALL2020}.  

In the present paper, we employ the \emph{local} version of \textsc{Gene}~\cite{Jenko_GENE}, which implements a standard flux-tube model~\cite{Beer1994}, that is, it considers only a sub-volume of the core plasma centered around a particular magnetic field line.
While these simulations retain full dependence of the magnetic geometry along the magnetic field line, the radial dependence of background quantities is determined from a first-order Taylor expansion.
Thus the flux-tube domains do not incorporate the full radial dependence of either the temperature and density profiles or the magnetic geometry.
However, they are well-suited to study local instability and turbulence phenomena in fusion plasmas, such as gyro-radius scale drift wave instabilities, due to the highly anisotropic nature of field-aligned fluctuations.
A mixture of finite differences and pseudo-spectral methods are employed to discretize the configuration space, while collisions are incorporated using finite volume methods.
The time evolution of the spatially discretized operators is performed using an adaptive fourth-order Runge-Kutta (RK4) time integrator.

\section{Learning higher-dimensional parametric data-driven reduced models using sparse grids}
\label{sec:proposed_approach}

This section details our approach for learning higher-dimensional parametric ROMs based on optDMD and sparse grid interpolation with $(L)$-Leja points.
Section~\ref{subsec:plasma_setup} summarizes the high-fidelity simulation setup, while Sec.~\ref{subsec:plasma_learning-parametric_ROMs} describes the construction of parametric ROMs using the aforementioned techniques for linear plasma micro-instability simulations.

\subsection{Setup for high-fidelity linear plasma micro-instability simulations}
\label{subsec:plasma_setup}

As a real-world example that can benefit from parametric reduced modeling, this paper focuses on constructing data-driven ROMs to quickly and effectively predict the full phase space structure of dominant micro-instability modes in fusion plasma scenarios with high-dimensional input parameter spaces.
Specifically, we consider linear, local, flux-tube gyrokinetic simulations with one plasma species (i.e., $n_s=1$).
In linear simulations, $n_{k_y}=1$.
These linear simulations are essential for identifying underlying micro-instabilities, such as ion or electron temperature gradient (ITG/ETG) driven modes, trapped electron modes (TEMs), and micro-tearing modes (MTMs), which form the foundation for nonlinear transport analysis in magnetically confined fusion plasmas.
Linear instability characteristics, such as linear growth rates, propagation frequencies, and eigenmode structures, provide critical information on the underlying sources of free energy that drive unwanted turbulent transport.
These instabilities have complicated dispersion relations that often require numerical simulation to understand, with nontrivial and often nonlinear dependencies on parameters such as wavelength, plasma pressure profiles, and magnetic field structure.
Furthermore, establishing their threshold values and transitions in parameter space is particularly useful, as it often informs the selection of input parameters for more complex nonlinear simulations.
To this aim, the ROM predictions will include the complex eigenvalue that provides the growth rate and frequency of the unstable, dominant eigenmode (the remaining eigenvalues correspond to stable modes when the ROM reduced dimension $r>1$) and the time evolution of the distribution function over a target time horizon  $[t_{i}, t_{f}]$ over the micro-instability saturated phase.
The latter can be used to derive other relevant quantities such as the electrostatic potential $\phi$, the parallel component of the vector potential $A_\parallel$, and the parallel component of the effective magnetic field $B^*_\parallel$ along the magnetic field direction $z$, defined in Sec.~\ref{subsec:background_gyrokinetic_modeling}.

The training data for constructing parametric data-driven ROMs will be generated using the linear initial-value solver provided by \textsc{Gene}.
In this approach, the nonlinear terms in Eq.~\eqref{eq0} are neglected, allowing the governing equations~\eqref{eq:FOM_general} to be compactly written as:
\begin{equation} \label{eq:FOM_linear_gyrokinetics}
    \frac{\mathrm{d}\mathbf{g}}{\mathrm{d} t}(t; \pmb{\uptheta}) = \mathbf{A}\mathbf{g}(t; \pmb{\uptheta}),
\end{equation}
where $\mathbf{A} \in \mathbb{C}^{n \times n}$ and $\mathbf{g}(t; \pmb{\uptheta}) \in \mathbb{C}^n$ denotes the semi-discrete form of the distribution function for the underlying plasma species.
The system is advanced in time using the RK4 time integrator, starting from an initial random distribution function.
After an initial transient phase, the growth rate of the distribution function will asymptote to that of the fastest growing instability. 
The simulation then terminates when the relative change in the running average of this growth rate drops below $10^{-3}$, a tolerance typically sufficient in practice.
In these simulations, the time domain of interest, $[t_i, t_f]$, is measured in units $L_{ref} / c_s$, where $L_{ref}$ is the reference length scale and $c_s$ is the ion sound speed.

We will also analyze the full spectrum of the reduced linear operators generated by our ROMs, particularly when the reduced dimension $r>1$.
To this aim, we can employ the \textsc{Gene} linear eigenvalue solver~\cite{Kammerer2008} to compute the spectrum for the reference data.
In this solver, an iterative Jacobi-Davidson solver with additive Schwarz preconditioning from the SLEPc library \cite{Hernandez2005, Romero2014} is applied to a matrix-free representation of the left side of Eq.~\ref{eq0}.
A shift parameter for shift-and-invert preconditioning may also be utilized to accelerate convergence in a particular region of the spectrum.
The stopping criterion for these eigenvalue calculations requires the eigenvalue residuals be less than $10^{-8}$ for each eigenvalue/eigenvector pair.
Using eigenvalue calculations to study the subdominant and stable eigenmode spectrum is critical to understanding the complex nonlinear dynamics of tokamaks \cite{Hatch2013, Whelan2018} and stellarators \cite{Faber2018, Pueschel2016}.

Two different instability scenarios are examined in this paper, one focusing on ITG modes and the other on ETG modes.
In both cases, simulations are performed by considering only gyrokinetic ions (electrons) for the ITG (ETG) parameter set, while the other plasma species are considered ``adiabatic''~\cite{Jenko_GENE}.
We note that while a single gyrokinetic species simulation inherently has limiting assumptions, it remains physically relevant.
The ITG is expected to be a major limiting factor in plasma performance in fusion scenarios, and has recently been shown to limit performance of the W7-X stellarator \cite{Beurskens_2021}.
Similarly, the ETG is expected to play a critical role in determining the performance of the tokamak pedestal~\cite{walker}, which is crucial to achieving high performance in tokamaks~\cite{Wagner1982}.
Both of these instabilities can be efficiently simulated by using only a single gyrokinetic species, which substantially reduces the cost required for the application presented here.
Nevertheless, the phase space states for these simulations are still complex and predicting the full 5D phase space is a nontrivial task.

One of our long-term goals is to extend this methodology to nonlinear gyrokinetic simulations of turbulent transport. 
This is, however, a highly nontrivial task from both computational and physical perspectives. 
On the computational side, training parametric ROMs will be challenging because the resulting simulation datasets are extremely large, necessitating parallel computing resources for efficient training~\cite{Fa25}. 
On the physics side, nonlinear gyrokinetic turbulence exhibits broadband spectra, intermittent transients, and energy transfer across scales via quadratic couplings (e.g., zonal flows), while also incurring substantially higher computational costs per parameter. 
In this context, we aim to develop parametric ROMs for the full distribution function where feasible, with the goal of capturing relevant statistics (e.g., time-averaged heat fluxes and fluctuation spectra) as well as coherent structures. 
This will be achieved through a combination of higher reduced dimensions, physics-based constraints, adaptive sparse-grid sampling to generate training data, and Operator Inference (OpInf)~\cite{KPW24,PW16} to model quadratic terms.

\subsection{Learning higher-dimensional parametric reduced models using $(L)$-Leja sparse grid interpolation} \label{subsec:plasma_learning-parametric_ROMs}

The strategy for selecting the $n_p$ training parameter instances, $\pmb{\uptheta}_1, \pmb{\uptheta}_2, \ldots, \pmb{\uptheta}_{n_p} \in \mathcal{D}$, is critical for the efficient construction of parametric ROMs and remains a significant open challenge in parametric model reduction.
In this work, we propose integrating sparse-grid interpolation based on $(L)$-Leja points into the reduced-order modeling procedure, as these points are well suited for high-dimensional parameter spaces and computationally expensive high-fidelity models.
For the specific example of plasma micro-instability simulations considered in the present paper (which are linear in the phase-space variables), we will construct the sparse-grid-accelerated ROMs based on optDMD.
However, the presented methodology is broadly applicable and can be extended to data-driven ROM approaches targeting nonlinear systems, such as OpInf.

Let $L$ denote a user-defined sparse-grid level associated with $N = n_p$ parameter instances. 
Based on our experience, relatively low grid levels (e.g., $L = 3$ or $4$) are typically sufficient when the target function to be interpolated is smooth. 
Because $(L)$-Leja sparse grids are nested, we can begin with a level-$1$ grid (containing a single point) and incrementally increase the level until the desired accuracy is achieved.

For each of the $n_p$ sparse grid training parameter instances $\pmb{\uptheta}_1, \pmb{\uptheta}_2, \ldots, \pmb{\uptheta}_{n_p} \in \mathcal{D}$, we perform the corresponding high-fidelity gyrokinetic simulation using \textsc{Gene} and collect the distribution function at $n_t$ time instants during the micro-instability saturated phase. 
Each 5D distribution function at a particular time instant is flattened into a column vector of size $n = n_{k_x} \times 1 \times n_{z}\times n_{v_\|}\times n_{\mu}$;  
subsequently, for each training parameter $\pmb{\uptheta}_j$, these $n_t$ snapshots are compiled into a complex-valued matrix $\mathbf{G}_j \in \mathbb{C}^{n \times n_t}$.

To remove any ambiguity in temporal alignment, we enforce a common-time policy for both training and predictions. 
Unless stated otherwise, we set $t_i\!=\!0$ and define $t_f$ to be the shortest available physical duration across all training runs.
All training snapshots are taken uniformly (after a fixed downsampling) on the same physical interval $t\in[0,\,t_f]$ with the same step $\Delta t$, yielding an identical snapshot count $n_t$ for every parameter instance.
Out-of-sample predictions are also evaluated on the same grid $t\in[0,\,t_f]$.

We next use $\mathbf{G}_1, \mathbf{G}_2, \cdots, \mathbf{G}_{n_p}$ to compute the global rank-$r$ POD basis.
This is done by first stacking the distribution function data horizontally into a snapshot matrix $\mathbf{G} = [\,\mathbf{G}_1\;\mathbf{G}_2\;\cdots\;\mathbf{G}_{n_p}] \in\mathbb{C}^{n\times n_pn_t}$ followed by computing its thin SVD,
\begin{equation*}
    \mathbf{G} = \mathbf{U} \mathbf{\Sigma} \mathbf{V}^*,
\end{equation*}
where $\mathbf{U} \in \mathbb{C}^{n \times n_p n_t}$ has unitary columns and contains the left singular vectors, $\bm{\Sigma} \in \mathbb{R}^{n_p n_t \times n_p n_t}$ is a diagonal matrix containing the singular values of $\mathbf{G}$ in non-decreasing order $\sigma_1 \geq \sigma_2 \geq \ldots \geq \sigma_{n_p n_t} \geq 0$, where $\sigma_j$ denotes the $j$th singular value, and $\mathbf{V} \in \mathbb{C}^{n_p n_t \times n_p n_t}$ has unitary rows and contains the right singular vectors.
The first $r \ll n$ columns of $\mathbf{U}$, i.e., the left singular vectors corresponding to the $r$ largest singular values, form the POD basis $\mathbf{U}_r \in \mathbb{C}^{n \times r}$.
The reduced dimension $r$ can be prescribed a-priori or chosen based on an energy criterion such as 
\begin{equation*} 
    \frac{\sum_{i=1}^{r} \sigma_i^2}{\sum_{i=1}^{n_p n_t} \sigma_i^2} \geq E_{\%},
\end{equation*}
where $E_{\%}$ is a user-specified energy truncation threshold (e.g., $E_{\%} = 99.99\%$).

Since one of the goals of the parametric ROMs in this paper is to predict the eigenvalue of the dominant micro-instability mode, we use the reduced basis $\mathbf{U}_r$ and optDMD to directly compute the spectra of the reduced operators $\widehat{\mathbf{A}}_1, \widehat{\mathbf{A}}_2, \ldots, \widehat{\mathbf{A}}_{n_p} \in \mathbb{C}^{r \times r}$, along with the associated optDMD modes and reduced initial conditions. 
This approach allows us to compute these quantities within the reduced coordinate system defined by $\mathbf{U}_r$.
If necessary, the reduced linear operators themselves can then be trivially reconstructed from their spectral decompositions.

The optDMD procedure is initialized with the continuous-time eigenvalues 
\begin{equation*}
\widehat{\pmb{\upalpha}}_j^{(0)} = \left[\, \widehat{\upalpha}^{(0)}_{1, j} \;  \widehat{\upalpha}^{(0)}_{2, j} \; \cdots \widehat{\upalpha}^{(0)}_{r, j} \,\right] \;=\; \left[\,  \frac{\ln\left(\widehat{\uplambda}_{1,j}^{(0)}\right)}{\Delta t}   \;  \frac{\ln\left(\widehat{\uplambda}_{2,j}^{(0)}\right)}{\Delta t} \; \cdots \; \frac{\ln\left(\widehat{\uplambda}_{r,j}^{(0)}\right)}{\Delta t} \,\right] \in \mathbb{C}^r,
\end{equation*}
where $\left\{\widehat{\uplambda}^{(0)}_{i, j}\right\}_{i=0}^r$ denote the discrete eigenvalues computed with standard DMD and $\Delta t$ is the uniform sampling interval. 
The optDMD algorithm computes the $r$ reduced continuous-time eigenvalues $\widehat{\pmb{\upalpha}}_j = \left[\, \widehat{\upalpha}_{1, j} \;  \widehat{\upalpha}_{2, j} \; \cdots \; \widehat{\upalpha}_{r, j} \,\right] \in \mathbb{C}^{r}$, $r$ optDMD modes $\widehat{\pmb{\Phi}}_j \in \mathbb{C}^{n\times r}$, and $r$ reduced amplitudes (i.e., initial conditions) $\widehat{\mathbf{b}}_j\in\mathbb{R}^r$ for $j=1, 2, \ldots, n_p$.
The eigenvectors of $\widehat{\mathbf{A}}_j$ are computed as $\widehat{\mathbf{W}}_{j} \;=\;\mathbf{U}_r^*\,\widehat{\pmb{\Phi}}_j \;\in\;\mathbb{C}^{r\times r}.$

To ensure consistent ordering of modes across parameter samples, we align each $(\widehat{\pmb{\upalpha}}_j, \widehat{\mathbf{W}}_{j}, \widehat{\mathbf{b}}_j)$ to a fixed reference case $j^\star$. 
Let $\widehat{\mathbf{W}}_{\mathrm{ref}}=\widehat{\mathbf{W}}_{j^\star}=[ w^{\mathrm{ref}}_1,\dots,w^{\mathrm{ref}}_r]$.
For each \(j\), we form the cosine-similarity matrix \cite{allemang1982correlation}
\begin{equation*}
\mathbf{S} \in \mathbb{R}^{r\times r}, \quad S_{k\ell}
=\frac{\bigl|\langle w^{\mathrm{ref}}_k,\ \widehat{w}_{j,\ell}\rangle\bigr|}
       {\lVert w^{\mathrm{ref}}_k\rVert_2\,\lVert \widehat{w}_{j,\ell}\rVert_2}
\in[0,1],\qquad k,\ell=1,\dots,r.
\end{equation*}
We then solve the linear assignment problem on $\mathbf{S} $ (via the Hungarian algorithm \cite{munkres1957algorithms}) to obtain a one-to-one matching that maximizes the total selected similarity.
The columns of \(\widehat{\mathbf{W}}_{j}\) and the entries of \(\widehat{\pmb{\upalpha}}_j\) and \(\widehat{\mathbf{b}}_j\) are then reordered accordingly.

In the next step, we use the aligned triplets $\{\widehat{\pmb{\upalpha}}_j, \widehat{\mathbf{W}}_{j}, \widehat{\mathbf{b}}_j\}_{j=1}^{n_p}$ to compute predictions for an arbitrary $\pmb{\uptheta}^\ddagger \in \mathcal{D}$ beyond training.
To this end, we interpolate the pairs $\{(\pmb{\uptheta}_j, \widehat{\pmb{\upalpha}}_j)\}_{j=1}^{n_p}$, $\{(\pmb{\uptheta}_j, \widehat{\mathbf{W}}_{j})\}_{j=1}^{n_p}$, and $\{(\pmb{\uptheta}_j, \widehat{\mathbf{b}}_j)\}_{j=1}^{n_p}$ using sparse grid Lagrange interpolation with $(L)$-Leja points as described in Sec.~\ref{subsec:background_SG_interp}.
For a given $\pmb{\uptheta}^\ddagger \in \mathcal{D}$, this yields 
$\{
\widetilde{\pmb{\upalpha}}(\pmb{\uptheta}^\ddagger),
\widetilde{\mathbf{W}}(\pmb{\uptheta}^\ddagger),
\widetilde{\mathbf{b}}(\pmb{\uptheta}^\ddagger)
\}$,
where $\widetilde{\pmb{\upalpha}}(\pmb{\uptheta}^\ddagger) = \mathcal{U}^d_{\mathcal{L}}[\hat{\pmb{\upalpha}}](\pmb{\uptheta}^\ddagger)$,
$\widetilde{\mathbf{W}}(\pmb{\uptheta}^\ddagger) = \mathcal{U}^d_{\mathcal{L}}[\hat{\mathbf{W}}](\pmb{\uptheta}^\ddagger)$,
$\widetilde{\mathbf{b}}(\pmb{\uptheta}^\ddagger) = \mathcal{U}^d_{\mathcal{L}}[\hat{\mathbf{b}}](\pmb{\uptheta}^\ddagger)\}$, and $\mathcal{L}$ denotes the multiindex set~\eqref{eq:std_sg_index_set} for the chosen sparse grid level $L$.
The first complex eigenvalue in $\widetilde{\pmb{\upalpha}}(\pmb{\uptheta}^\ddagger)$ provides the growth rate and frequency of the dominant eigenmode; when $r>1$, the remaining eigenvalues correspond to stable micro-instability modes.
The interpolated data $\{
\widetilde{\pmb{\upalpha}}(\pmb{\uptheta}^\ddagger),
\widetilde{\mathbf{W}}(\pmb{\uptheta}^\ddagger),
\widetilde{\mathbf{b}}(\pmb{\uptheta}^\ddagger)
\}$ 
can also be used to predict the time evolution of the distribution function $\mathbf{g}(t; \pmb{\uptheta}^\ddagger)$ over the target time horizon $[0, t_f]$. 
More specifically, we can compute $\widetilde{\mathbf{g}}(t; \pmb{\uptheta}^\ddagger) \approx \mathbf{g}(t; \pmb{\uptheta}^\ddagger)$ at a time instant $t \in [0, t_f]$ as 
\begin{equation*}
\mathbf{g}(t; \pmb{\uptheta}^\ddagger) \approx \widetilde{\mathbf{g}}(t; \pmb{\uptheta}^\ddagger)
        = \widetilde{\pmb{\Phi}}(\pmb{\uptheta}^\ddagger)\,\exp{(\mathrm{diag}(\widetilde{\pmb{\upalpha}}(\pmb{\uptheta}^\ddagger)) t)}\,\widetilde{\mathbf{b}}(\pmb{\uptheta}^\ddagger)
        \;\in\;\mathbb{C}^n,   
\end{equation*}
where $\widetilde{\pmb{\Phi}}(\pmb{\uptheta}^\ddagger) = \mathbf{U}_r\,\widetilde{\mathbf{W}}(\pmb{\uptheta}^\ddagger) \;\in\;\mathbb{C}^{n\times r}$ denotes the full-space DMD modes.
The ROM predicted distribution function can be subsequently used to derive relevant physics quantities (see Sec.~\ref{sec:results}).

We note that interpolating eigenpairs to compute predictions for arbitrary parameter instances was also considered in Ref.~\cite{Hu23}. 
However, Ref.~\cite{Hu23} used standard DMD to obtain $\hat{\mathbf{A}}$ followed by its eigendecomposition, whereas here we employ optDMD to directly compute the eigenpairs, as discussed above. 
Furthermore, Ref.~\cite{Hu23} used nearest neighbors to select parameter instances for interpolation, which is affected by the curse of dimensionality, while our approach mitigates this via sparse-grid interpolation, making it suitable for higher-dimensional parametric scenarios.
Finally, we use a reduced global basis instead of local bases as in Ref.~\cite{Hu23}. 
Using local bases is possible in our framework as well (e.g., transporting subspaces along Grassmann geodesics to interpolate modes), but this introduces additional alignment/transport steps and can suffer from mode crossings and discontinuities. 
In general, in micro-instability scenarios with a single dominant mode, such as those considered in the present paper, a global basis constructed from data at a set of representative parameter values is generally sufficiently expressive to capture the system behavior for any parameter instance within the prescribed bounds.
Our code is publicly available at \url{https://github.com/kevinsinghgill/parametric_ROMs} and is suitable for parametric data-driven reduced modeling beyond linear plasma micro-instability simulations.

\begin{remark} \label{remark:1D_parameter_spaces}
Sparse grid approximations are designed for problems with at least two parameters, i.e., $d \geq 2$.
When dealing with a single input parameter, $\uptheta \in \mathcal{D} \subseteq \mathbb{R}$, we can generate training instances $\uptheta_1, \uptheta_2, \ldots, \uptheta_{n_p}$ using standard strategies such as a uniform grid or pseudo-random sampling. 
Interpolation can then be performed using 1D methods such as spline interpolation.
\end{remark}

\begin{remark} \label{remark:SG_limitations}
While the cardinality of $(L)$-Leja sparse grids grows slowly with the number of parameters and grid level, it can still become prohibitively large for computationally expensive problems when the dimensionality (d) or sparse grid level (L) are large. 
One approach to mitigate this is adaptivity, which has proven efficient in high-dimensional problems~\cite{FDJ21}. 
We plan to integrate adaptivity into our parametric reduced modeling procedure in future research.
\end{remark}

\begin{remark} \label{remark:extrapolation}
The proposed parametric reduced-order modeling workflow is intended for predictions in the interpolative regime, i.e., for parameter instances within $\mathcal{D}$. 
Because our approach employs sparse-grid interpolation with Lagrange polynomials, it is expected to perform poorly in extrapolative settings.
Nevertheless, previous work on surrogate models for heat fluxes~\cite{FMJ24} suggests that regression-based approaches could enable robust extrapolation in cases dominated by a single micro-instability mode, as considered here.
Extending the present framework to support reliable extrapolation is left for future research.
\end{remark}

\section{Numerical experiments in two plasma micro-instability simulation scenarios} \label{sec:results}

This section presents our numerical results in two plasma micro-instability simulation scenarios.
Section~\ref{subsec:res_CBC} examines the Cyclone Base Case (CBC) benchmark, characterized by ITG micro-instability modes. 
Here, we assess the predictive capabilities of optDMD ROMs in two ways: extrapolating predictions beyond a training time horizon for fixed input parameters, and for variations in one plasma parameter, the binormal wave number, $k_y \rho_s$.
Section~\ref{subsec:res_ETG} considers a real-world scenario: ETG-driven micro-instability in the pedestal region of a tokamak. 
We evaluate the efficiency and predictive power of sparse-grid-accelerated parametric optDMD ROMs under variations in six key plasma parameters.
To our knowledge, this represents one of the first investigations into parametric reduced modeling for gyrokinetic simulations of plasma micro-instability. 

Both the high-fidelity and reduced modeling computations were performed on the Perlmutter supercomputer\footnote{\url{https://docs.nersc.gov/systems/perlmutter/architecture/}} at the National Energy Research Scientific Computing Center (NERSC).
Each CPU and GPU node is equipped with two AMD EPYC 7763 (Milan) CPUs with 64 cores per CPU and 512 Gb DDR4 memory total.
The high-fidelity \textsc{Gene} simulations to generate training and reference data for our ROMs were performed on a single Perlmutter compute node utilizing four NVIDIA A100 GPUs with 80 Gb on-card memory each, linked via GPU-aware MPI.
All reduced modeling computations and \textsc{Gene} eigenvalue calculations were performed on one Perlmutter CPU node.

\subsection{Cyclone Base Case benchmark with deuterium ions and ion temperature gradient driven modes}
\label{subsec:res_CBC}

We begin by examining the CBC benchmark scenario presented in~\cite{Di00}.
We consider a setup with a single plasma species, namely deuterium ions, characterized by ITG micro-instabilities.
The electrons are assumed to be adiabatic. 
For the magnetic geometry, this setup utilizes an $s$--$\alpha$ circular model.
The high-fidelity model is discretized using a grid with $n = n_{k_x}\times 1\times n_{z}\times n_{v_\|}\times n_{\mu}=15\times1\times16\times32\times8 = 61,440$ degrees of freedom in the five-dimensional position--velocity phase space.
Throughout all experiments, the input plasma parameters are set to the default nominal values defined for this benchmark scenario.

\subsubsection{Reduced-order modeling predictions beyond a training time horizon for fixed parameter values}
\paragraph{Setup for reduced-order modeling}
To develop initial intuition about data-driven ROMs in micro-instability simulations, we begin by investigating predictive optDMD ROMs and their ability to predict plasma dynamics beyond the training time horizon under a fixed set of input parameters.
We furthermore compare the prediction capabilities of standard DMD and optDMD.
Predictive ROMs are particularly useful for computing long-term predictions that remain computationally too expensive in terms of the high-fidelity simulation model. 
Given data over a training horizon [$0, t_t$], these ROMs predict the time evolution of the distribution function over [$t_t, t_f$], where $t_t < t_f$.
Here, we focus on a CBC simulation with $k_y \rho_s = 0.30$, corresponding to ITG-driven micro-instability.

The high-fidelity \textsc{Gene} initial-value simulation is performed over the time domain $[0, 32.993]$ measured in units $R / c_s$, where $R$ is the major radius of the toroidal plasma.
The \textsc{Gene} runtime is $125.154$ seconds using four GPUs on a single GPU node on Perlmutter.
We save the complex-valued distribution function to disk every $100$ time iterations.
We explore two setups for constructing linear ROMs for forecasting beyond training.
In both setups, we use approximately $40\%$ of the simulation time history for training.
In the first setup, the training begins after the transient phase at $t_i = 6.6$.
The training horizon ends at $t_t = 19.801$, yielding $n_t = 132$ training snapshots.
Therefore, $\mathbf{G} \in \mathbb{C}^{ 61,440 \times 132}$.
This is generally the preferred approach in practice. 
Due to the transient nature of subdominant plasma micro-instabilities, practitioners are usually interested only in final dominant micro-instability. 
For a more comprehensive analysis, we also consider the more challenging setup in which training begins at $t_i = 0$, thereby including the transient phase, and ends at $t_t =13.201$.
With this setup, we have $n_t = 199$ snapshots for training, which means that $\mathbf{G} \in \mathbb{C}^{ 61,440 \times 199}$.

\begin{figure}[htbp!]
  \centering 
  \includegraphics[width=1.0\textwidth] {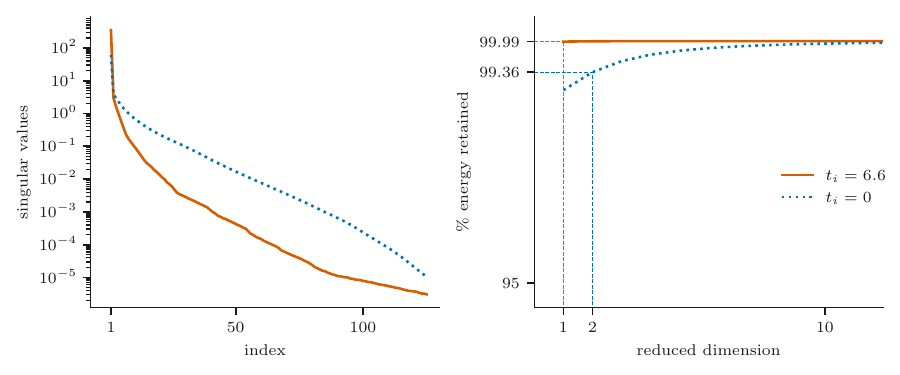}
  \caption{Cyclone Base Case benchmark scenario: Singular values (left) and corresponding retained energy (right) when constructing an optDMD ROM for predictions beyond the training horizon. The orange curves represent the case in which the training horizon begins at $t_i = 6.6$, after the transient phase, whereas the blue curves correspond to a training horizon starting at $t_i = 0$, which includes the transient phase.} 
  \label{fig:CBC-timepred-svals}
\end{figure}

\paragraph{Reduced-order modeling predictions beyond the training horizon}
Figure~\ref{fig:CBC-timepred-svals} plots the singular values (left) and the corresponding retained energy (right) for both considered setups. 
As expected for a simulation dominated by a single micro-instability mode, the first mode captures the majority of the energy in both cases. 
Furthermore, the singular values decay rapidly.
For the first setup, over $99\%$ of the total energy is contained within the first mode, while for the second setup, this is achieved by the first two modes. 
Consequently, we construct optDMD ROMs with reduced dimensions $r=1$ and $r=2$, respectively.

We begin by quantitatively assessing the prediction accuracy achieved by the optDMD ROMs for both scenarios under consideration. 
To obtain a more comprehensive picture, we also examine how the accuracy varies with the size of the training window. 
In particular, we reduce the amount of training data from $40\%$ of the simulation history to $20\%$ and $10\%$.
Table~\ref{tab:cbc_transient_window_scan} lists the corresponding training time horizons and the relative errors in the real and imaginary components of the full distribution function over the prediction horizon, as well as the relative errors in the real and imaginary parts of the eigenvalue associated with the dominant ITG mode. 
The reference eigenvalue computed with \textsc{Gene} is $\upalpha_{\mathrm{ref}} = 0.264 + 0.790\mathrm{i}$.
The upper half of the table reports results for training initiated at $t_i = 6.6$ with a fixed ROM dimension of $r = 1$. 
In this regime, the errors decrease monotonically as the training window expands, indicating that the ROM accuracy is not strongly sensitive to the window size when the model is trained in the quasi--steady-state phase of the dynamics.
The lower half of Table~\ref{tab:cbc_transient_window_scan} presents results for training initiated at $t = 0.0$ with a fixed ROM dimension of $r = 2$. 
Here, the same monotonic trend is observed; however, when the training window is short and captures only the transient portion of the dynamics, the resulting errors are substantially larger. 
As the training window increases and the dataset begins to include quasi--steady-state behavior, these errors steadily decrease--falling below $7.5\%$ for the distribution function and $0.35\%$ for the growth rate and frequency of the dominant eigenmode--levels that are acceptable given the increased difficulty of this scenario.

\begin{table}[htbp!]
\centering
\begin{tabular}{{|c|c|c|c|c|c|}}
\hline
\begin{tabular}{@{}c@{}}[$t_i, t_t$]\end{tabular} &
\begin{tabular}{@{}c@{}}Re($\mathbf{g}$) error (\%)\end{tabular} & \begin{tabular}{@{}c@{}}Im($\mathbf{g}$) error (\%)\end{tabular} &
\(\tilde{\upalpha}_{\mathrm{ROM}}\) &
\begin{tabular}{@{}c@{}}Growth rate\\error (\%)\end{tabular} &
\begin{tabular}{@{}c@{}}Frequency\\error (\%)\end{tabular} \\
\hline
$[6.600, 9.900]$ & 11.165 & 10.890 & 0.259 + 0.790\(\mathrm{i}\)  & 1.876 & 0.031 \\
$[6.600, 13.202]$ & 4.012 & 4.515 & 0.262 + 0.789\(\mathrm{i}\)  & 0.817 & 0.184 \\
$[6.600, 19.801]$ & 0.798 & 0.870 & 0.263 + 0.790\(\mathrm{i}\) & 0.343 & 0.003 \\
\hline
$[0.000, 3.300]$  & 100.980 &  100.243 & 0.140 + 0.859\(\mathrm{i}\) & 47.130 & 8.689 \\
$[0.000, 6.600]$  & 63.455 & 70.004 & 0.278 + 0.803\(\mathrm{i}\)  & 5.392 & 1.704 \\
$[0.000, 13.202]$ & 7.301 & 5.398 & 0.265 + 0.788\(\mathrm{i}\)  & 0.341 & 0.312 \\
\hline
\end{tabular}
\caption{Cyclone Base Case benchmark scenario: Effect of training window size on the predicted distribution function and dominant ITG mode.}
\label{tab:cbc_transient_window_scan}
\end{table}

\begin{figure}[htbp!]
  \centering 
  \includegraphics[width=1.0\textwidth]{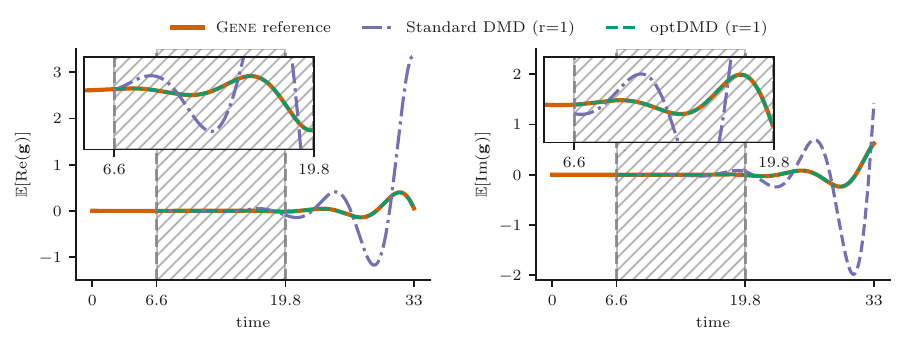}
  \caption{Cyclone Base Case benchmark scenario:  Comparison between the expected value of the real (left) and imaginary (right) components of the distribution function computed using standard DMD and optDMD with reduced dimension $r=1$. The training horizon $[6.6, 19.8]$ starts after the transient phase. 
  } 
  \label{fig:CBC_dmd_optdmd}
\end{figure}

\begin{figure}[htbp!]
  \centering 
  \includegraphics[width=1.0\textwidth]{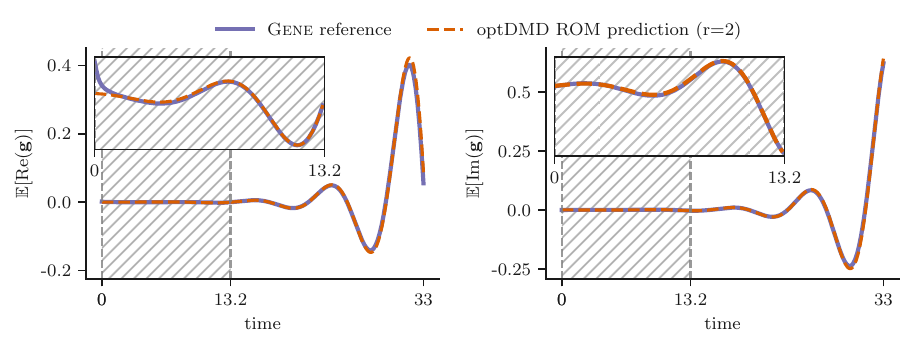}
  \caption{Cyclone Base Case benchmark scenario: The expected value of the real (left) and imaginary (right) components of the distribution function obtained using the  $r=2$ optDMD ROM versus the corresponding reference \textsc{Gene} data over the target time horizon.
  The training horizon $[0.0, 13.2]$ includes the transient phase.} 
  \label{fig:CBC-time-prediction-inset}
\end{figure}

To provide additional insights into the predictions of the optDMD ROM, Figs.~\ref{fig:CBC_dmd_optdmd} and \ref{fig:CBC-time-prediction-inset} compare the expectations of the real (left) and imaginary (right) components of the distribution function over their respective time horizons with the \textsc{Gene} reference data, when using $40\%$ of the time history for training. 

We also perform a comparison between standard DMD and optDMD.
Figure~\ref{fig:CBC_dmd_optdmd} shows the standard and optDMD comparison when the training window begins after the transient regime ($t_i = 6.6$) using a reduced dimension of $r = 1$.
Standard DMD (dashed magenta line) exhibits noticeable deviations from the reference solution, particularly near 
the end of the training interval, and these discrepancies amplify over the prediction 
horizon. 
In contrast, optDMD (dashed green line) maintains close agreement with the reference data across both 
the training and prediction intervals, illustrating its superior predictive performance in 
this setting.
Despite these accuracy differences, standard DMD yields an accurate estimate of the eigenvalue associated with the dominant eigenmode, consistent with observations reported in the literature (e.g., \cite{Du25}). 
Specifically, standard DMD produces \(\upalpha_{\mathrm{ROM}} = 0.263 + 0.789\,\mathrm{i}\), while optDMD 
yields \(\upalpha_{\mathrm{ROM}} = 0.263 + 0.790\,\mathrm{i}\). 

Figure~\ref{fig:CBC-time-prediction-inset} compares the expectation results obtained with optDMD using a reduced dimension of $r = 2$ against the reference \textsc{Gene} data for the more challenging scenario in which training begins at $t_i = 0$ and the first $40\%$ of the time history is used for training. 
The expectations computed from the ROM solutions align closely with the reference data, consistent with the results presented in Table~\ref{tab:cbc_transient_window_scan}; the minor discrepancies observed near the start of the time domain can be attributed to the complex transient dynamics. 

Finally, we report both the offline ROM construction costs and the online evaluation costs when using $40\%$ of the data for training.
For the first setup (training during the saturated phase), constructing the ROM required $2.391$ seconds on a single CPU core on Perlmutter. 
Evaluating this optDMD ROM over the target time horizon $[6.6, 32.993]$ took a total of $0.065$ seconds, with $0.033$ seconds spent over the training horizon and $0.032$ seconds over the prediction horizon.
Similarly, constructing the ROM for the second setup (training from $t_i = 0$) required $2.455$ seconds. 
Its evaluation over the target time horizon $[0, 32.993]$ took $0.080$ seconds, split between $0.033$ seconds for training and $0.047$ seconds for prediction.
These results therefore demonstrate a significant reduction in runtime for both considered setups. 
Our optDMD ROMs reduce the evaluation time from $125.154$~s (using four GPUs on a single GPU node on Perlmutter) to just $0.065$~s and $0.080$~s on a single CPU core, corresponding to speedup factors of $1,925$ and $1,564$, respectively.

\subsubsection{Reduced-order modeling predictions for variations in the normalized binormal wavenumber $k_y \rho_s$}
\label{subsubsec:CBC_2D_slices}
We next investigate the ability of the optDMD ROMs to predict the full state as the normalized binormal wavenumber, $k_y \rho_s$, is varied. 
We consider $k_y \rho_s$ values between $0.05$ and $0.55$, a range in which the ITG mode is destabilized. 
Since the parametric dimension is $d=1$, we do not employ $(L)$-Leja sparse grids for generating training data (sparse grids are defined for $d \ge 2$). 
Similarly, $(L)$-Leja-based Lagrange interpolation is not used for ROM predictions (cf.~Remark~\ref{remark:1D_parameter_spaces}). 

\paragraph{Training a parametric reduced model using data for $n_p=4$ parameter instances}
We first consider a training dataset comprising only $n_p=4$ instances of $k_y \rho_s$ including the extrema, namely $k_y \rho_s \in \{0.05, 0.20, 0.40, 0.55\}$.
Our objective is to evaluate the accuracy of a parametric ROM developed from this sparse set of training samples.
Details about the corresponding high-fidelity \textsc{Gene} runtimes for each of the four $k_y \rho_s$ values can be found in Table~\ref{tab:cbc-4training-details}.
All four training cases are simulated over the same time horizon, $t \in [0, 33.993]$, where $t_f = 33.993$ extends into the saturated phase of the micro-instability for all cases.
As in the previous experiment, we save the distribution functions to disk every $100$ time iterations. 
This amounts to a total of $n_t = 332$ snapshots per $k_y \rho_s$ instance, so the corresponding global snapshot matrix has size $61,440 \times 4\cdot332 = 61,440 \times 1,328$.
\begin{table}[htbp!]
\centering
\begin{tabular}{|c|c|c|}
\hline
	 \(k_y \rho_s\)  & \textsc{Gene} runtime (s) \\
\hline
	0.05 & 454.647   \\
    0.20 & 142.372   \\
    0.40 & 145.320   \\
    0.55 & 370.034   \\
\hline
\end{tabular}
\caption{Cyclone Base Case benchmark scenario: setup for parametric reduced modeling using $n_p = 4$ training instances.}
\label{tab:cbc-4training-details}
\end{table}

\begin{figure}[htbp!]
  \includegraphics[width=1.0\textwidth] {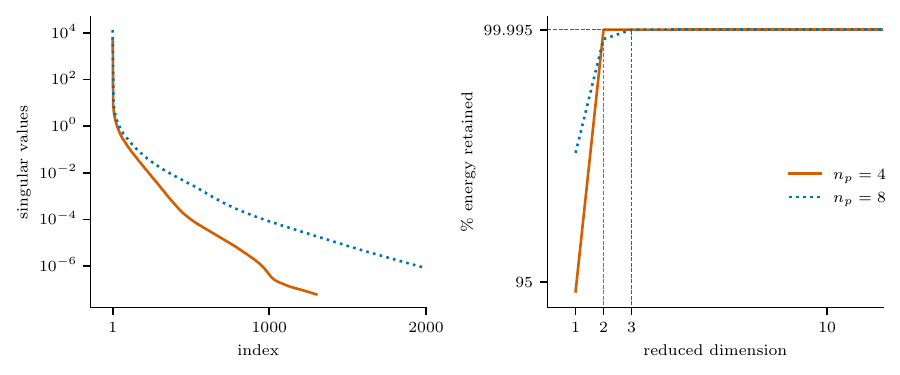}
  \caption{Cyclone Base Case benchmark scenario: Singular values (left) and corresponding retained energy (right) when constructing a parametric optDMD ROM using $n_p=4$ training parameter instances, namely $k_y \rho_s \in \{0.05, 0.20, 0.40, 0.55\}$ compared to using $n_p=8$ instances, i.e., $k_y \rho_s \in \{0.05, 0.12, 0.20, 0.28, 0.38, 0.40, 0.48, 0.55\}$.} 
  \label{fig:CBC-parametic-svals}
\end{figure}

Figure~\ref{fig:CBC-parametic-svals} plots the singular values on the left and the corresponding percentage of the retained energy on the right in solid orange line.
The singular values decay fast: only $r=2$ modes are sufficient to retain more than $99.99\%$ of the energy.
We therefore construct a parametric ROM with reduced dimension $r=2$.

For evaluating the parametric ROM's ability to predict beyond its training data, we consider six testing parameter instances $k_y \rho_s \in \{0.10, 0.15, 0.25, 0.30, 0.45, 0.50\}$. 
All six instances lie within the training interval $k_y\rho_s\in[0.05,\,0.55]$, placing them in the interpolative regime.
We do not investigate extrapolation outside this range; such analysis is left for future work.
Cubic spline interpolation is used to compute the corresponding optDMD ROM predictions.

Table~\ref{tab:cbc_parametric_ky_eig_4training} compares the eigenvalues corresponding to the dominant eigenmode obtained using our optDMD ROM (i.e., $\tilde{{\upalpha}}_{\mathrm{ROM}}$) against the reference eigenvalues computed with \textsc{Gene} (i.e., $\upalpha_{\mathrm{ref}}$). 
With the exception of $k_y \rho_s \in \{ 0.10, 0.15 \}$, where the growth rate relative errors are $25.602\%$ and $8.025\%$, all other errors remain $\lesssim 3\%$, thus indicating a reasonable prediction capability given that only $n_p=4$ training instances have been used.
The outlier at $k_y \rho_s=0.10$ can be attributed to the following factors.
First, only four non-uniformly spaced data points are available to compute predictions via cubic spline interpolation. 
Second, the reference growth rate at $k_y \rho_s=0.10$ is small in magnitude, which inflates relative errors even when the absolute error is modest.
In addition, from a physical standpoint, simulations at lower $k_y \rho_s$ (i.e., larger structures that take longer to resolve) are more susceptible to larger errors in the growth-rate and frequency calculations.

\begin{table}[htbp!]
\centering
\begin{tabular}{|c|c|c|c|c|}
\hline
	 &  \(\upalpha_{\mathrm{ref}}\) &\ $\tilde{{\upalpha}}_{\mathrm{ROM}}$  & Growth rate error (\%) & Frequency error (\%) \\
\hline
	0.10 & 0.084 + 0.219$\mathrm{i}$ &  0.106 + 0.221$\mathrm{i}$ & 25.602 & 1.092 \\
    0.15 & 0.152 + 0.353$\mathrm{i}$ &  0.164 + 0.356$\mathrm{i}$ & 8.025 & 0.895 \\
    0.25 & 0.248 + 0.644$\mathrm{i}$ &  0.241 + 0.640$\mathrm{i}$ & 2.938 & 0.603 \\
    0.30 & 0.264 + 0.790$\mathrm{i}$ &  0.256 + 0.784$\mathrm{i}$ & 3.122 & 0.761 \\
    0.45 & 0.190 + 1.185$\mathrm{i}$ &  0.191 + 1.191$\mathrm{i}$ & 0.718 & 0.526 \\
    0.50 & 0.130 + 1.301$\mathrm{i}$ &  0.129 + 1.310$\mathrm{i}$ & 0.953 & 0.691 \\
\hline
\end{tabular}
\caption{Cyclone Base Case benchmark scenario: Comparison between the ROM prediction ($r=2$, trained with $n_p=4$ instances) and the \textsc{Gene} reference result for the eigenvalue corresponding to the dominant ITG micro-instability mode, shown for six parameter instances beyond training.}
\label{tab:cbc_parametric_ky_eig_4training}
\end{table}

\paragraph{Increasing the number of training parameter instances to $n_p=8$}
We next investigate the effect of increasing the number of training parameter instances.
Specifically, we double the number of instances to $n_p = 8$, using $k_y \rho_s = \{0.05, 0.12, 0.20, 0.28, 0.38, 0.40, 0.48, 0.55\}$.
Details about the \textsc{Gene} high-fidelity simulations for the eight training parameter instances are found in Table~\ref{tab:cbc-8training-details}.
As before, we use the common physical interval $t\in[0, 33.993]$ and save the corresponding \textsc{Gene} distribution functions to disk every $100$ time instants, resulting in $n_t = 332$ snapshots per $k_y \rho_s$ instance.
Therefore, the global snapshot matrix $\mathbf{G}$ is of size $61,440 \times 8\cdot332 = 61,440 \times 2,656$.
\begin{table}[htbp!]
\centering
\begin{tabular}{|c|c|c|}
\hline
	 \(k_y \rho_s\)  & \textsc{Gene} runtime (s) \\
\hline
    0.05 & 454.647    \\
    0.12 & 202.679    \\
    0.20 & 142.372    \\
    0.28 & 124.049    \\
    0.38 & 135.065    \\
    0.40 & 145.320    \\
    0.48 & 193.450    \\
    0.55 & 370.034    \\
\hline
\end{tabular}
\caption{Cyclone Base Case benchmark scenario: setup for parametric reduced modeling using $n_p = 8$ training instances.}
\label{tab:cbc-8training-details}
\end{table}

Figure~\ref{fig:CBC-parametic-svals} plots the singular values on the left and the percentage of retained energy on the right using dashed blue lines.
The singular values exhibit a fast decay: only $r=3$ POD modes are sufficient to retain $99.99\%$ of the total energy. 
Consequently, we develop a parametric ROM with reduced dimension $r=3$.

We begin by assessing the optDMD ROM's capability to predict the growth rate and frequency of the dominant eigenmode. 
Table~\ref{tab:cbc_parametric_ky_eig_8training} presents a comparison between the ROM predictions and reference \textsc{Gene} results in terms of relative errors. 
Doubling the number of training parameters significantly improves the accuracy of the optDMD ROM predictions: the relative errors for $k_y \rho_s \in \{ 0.15, 0.25, 0.30, 0.45, 0.50 \}$ fall below $1\%$, while the growth rate error for $k_y \rho_s =0.10$ decreased from $25.602\%$ to $2.389\%$.
\begin{table}[htbp!]
\centering
\begin{tabular}{|c|c|c|c|c|}
\hline
	 &  \(\upalpha_{\mathrm{ref}}\) &\ $\tilde{{\upalpha}}_{\mathrm{ROM}}$  & Growth rate error (\%) & Frequency error (\%) \\
\hline
	0.10 & 0.084 + 0.219$\mathrm{i}$ &  0.086 + 0.216$\mathrm{i}$ & 2.389 & 1.549 \\
    0.15 & 0.152 + 0.353$\mathrm{i}$ &  0.151 + 0.354$\mathrm{i}$ & 0.760 & 0.395 \\
    0.25 & 0.248 + 0.644$\mathrm{i}$ &  0.248 + 0.644$\mathrm{i}$ & 0.014 & 0.001 \\
    0.30 & 0.264 + 0.790$\mathrm{i}$ &  0.264 + 0.790$\mathrm{i}$ & 0.157 & 0.041 \\
    0.45 & 0.190 + 1.185$\mathrm{i}$ &  0.190 + 1.185$\mathrm{i}$ & 0.147 & 0.032 \\
    0.50 & 0.130 + 1.301$\mathrm{i}$ &  0.129 + 1.301$\mathrm{i}$ & 0.720 & 0.032 \\
\hline
\end{tabular}
\caption{Cyclone Base Case benchmark scenario: Comparison between the ROM prediction ($r=3$, trained with $n_p=8$ instances) and the \textsc{Gene} reference result for the eigenvalue corresponding to the dominant ITG micro-instability mode, shown for six parameter instances beyond training.}
\label{tab:cbc_parametric_ky_eig_8training}
\end{table}

We next use the optDMD ROM to predict the full distribution function for the six testing instances, $k_y \rho_s \in \{0.10, 0.15, 0.25, 0.30, 0.45, 0.50\}$. 
This illustrates a key advantage of ROMs over surrogate models that predict only scalar quantities of interest, such as growth rates (e.g., Refs.~\cite{FDJ21, Fr23, Ko22}): ROMs can predict the full distribution function, offering substantially more information.
Table~\ref{tab:cbc-runtime-comparison-8training} presents the runtimes for the reference \textsc{Gene} simulations, performed using four GPUs on one Perlmutter GPU node over the full time horizons starting from $t=0$, and the online ROM evaluation costs over the time horizon of interest.
We first note the longer \textsc{Gene} runtimes for the extrema of $k_y \rho_s$, which is due to needing more iterations to reach convergence.
The offline cost for constructing the ROM with a reduced dimension $r=3$ is $40.640$ seconds. 
Online, evaluating the ROMs over the target time horizon costs at most $0.044$ seconds, achieving a speedup factor of three orders of magnitude compared to the high-fidelity \textsc{Gene} runtimes.
\begin{table}[htbp!]
\centering
\begin{tabular}{|c|c|c|c|}
\hline
	 \(k_y \rho_s\) 
 & \textsc{Gene} runtime (s)
 & ROM runtime (s) & Speedup factor \\
\hline
	  0.10 & 245.007 & 0.035 &  7,000  \\ 
    0.15 & 180.639 & 0.044 &  4,105 \\
    0.25 & 133.659 & 0.038 &  3,517 \\
    0.30 & 125.154 & 0.042 &  2,979 \\
    0.45 & 161.607 & 0.035 &  4,617 \\
    0.50 & 215.609 & 0.044 &  4,900 \\
\hline
\end{tabular}
\caption{Cyclone Base Case benchmark scenario: Comparison between the reference \textsc{Gene} and ROM runtimes ($r=3$, trained using $n_p=8$ parameter instances). The ROM offline construction cost is $40.640$ seconds.}
\label{tab:cbc-runtime-comparison-8training}
\end{table}

For parametric-variation predictions, computing errors between the full ROM and the reference distribution functions does not provide a reliable assessment of ROM accuracy. 
In the reference simulations performed with \textsc{Gene}, the distribution function may be rescaled (for example, when its amplitude exceeds an internal threshold), so a direct comparison between the ROM and reference solutions is not always meaningful.
More broadly, one of our goals is to develop ROMs for plasma micro-turbulence, where the nonlinear and chaotic nature of the dynamics renders classical error measures of limited practical value. 
It is therefore more informative to evaluate ROM performance using physically relevant diagnostics.
In the experiments presented here, we assess the ROM at the structure level through time-averaged two-dimensional projections of $\mathbf{g}$. 
In the following section, we further perform physics-based validation by comparing the electromagnetic fields derived from $\mathbf{g}$.

To compare the spatio-velocity structure predicted by the ROM against \textsc{Gene} in an interpretable way, we summarize the distribution function using two-dimensional averages as follows.
To simplify the notation, let $\mathbf{g}$ denote the distribution function (computed with the ROM or \textsc{Gene}) and let $n_{\mathrm{qs}}$ be the number of snapshots at the end of the time horizon of interest used for averaging (typically the last $100$–$200$ snapshots). 
With uniform sampling step $\Delta t$ and total count $n_t$, we define the average over the last $n_{\mathrm{qs}}$ snapshots as:
\begin{equation*}
\overline{\mathbf{g}}(k_x,z,v_{\parallel},\mu)
= \frac{1}{n_{\mathrm{qs}}}
   \sum_{m=n_t-n_{\mathrm{qs}}+1}^{n_t}
   \mathbf{g}\!\big(t_m,k_x,z,v_{\parallel},\mu\big),
\qquad
t_m := t_i + (m-1)\,\Delta t .
\end{equation*}

Time averaging suppresses fast oscillations and random phase shifts, highlighting persistent patterns relevant to transport. 
Afterwards, for each of the six possible two-dimensional combinations of directions, we compute 2D representations by averaging over the other two dimensions. 
For example, to obtain the projection onto the $(k_x, z)$ plane, we average over $v_{\parallel}$ and $\mu$:
\begin{equation*}
    \langle \bar{\mathbf{g}} \rangle_{k_x, z}(k_x, z) = \frac{1}{n_{v_{\parallel}} n_{\mu}} \sum_{v_{\parallel}} \sum_{\mu} \bar{\mathbf{g}}(k_x, z, v_{\parallel}, \mu)
\end{equation*}
where $n_{v_{\parallel}}$ and $n_{\mu}$ denote the number of grid points along the averaged dimensions. 
This process is repeated for all six possible combinations. 
Each two-dimensional projection highlights different aspects of the underlying physics.
For example, the $(k_x, z)$ projection reveals the relationship between radial scale and magnetic-field-line localization, while the $(v_{\parallel}, \mu)$ projection captures the anisotropy in velocity space. 
Despite their distinct interpretations, all projections indicate where the fluctuation content is concentrated in the structure of the dominant instability.

For visualization purposes, we separate the real and imaginary parts of the 2D averages. 
Each resulting pair of 2D slices is then normalized by its respective maximum absolute value. 
This procedure scales each pair to the range $[-1, 1]$, enabling a standardized visual comparison of the structure and relative magnitude of the data. 
The final normalized quantities used for plotting, denoted by $\tilde{\mathbf{g}}$, are therefore given by:
\begin{align*}
    \mathrm{Re}[\tilde{\mathbf{g}}_{i,j}] = \frac{\mathrm{Re}[\langle \bar{\mathbf{g}} \rangle_{i,j}]}{\max |\mathrm{Re}[\langle \bar{\mathbf{g}} \rangle_{i,j}]|}, \qquad
    \mathrm{Im}[\tilde{\mathbf{g}}_{i,j}] = \frac{\mathrm{Im}[\langle \bar{\mathbf{g}} \rangle_{i,j}]}{\max |\mathrm{Im}[\langle \bar{\mathbf{g}} \rangle_{i,j}]|},
\end{align*}
where $(i,j) \in \{k_x, z, v_{\parallel}, \mu\}^2$ and $i \neq j$.

\begin{figure}[htbp!]
  \centering 
  \includegraphics[width=1.0\textwidth]{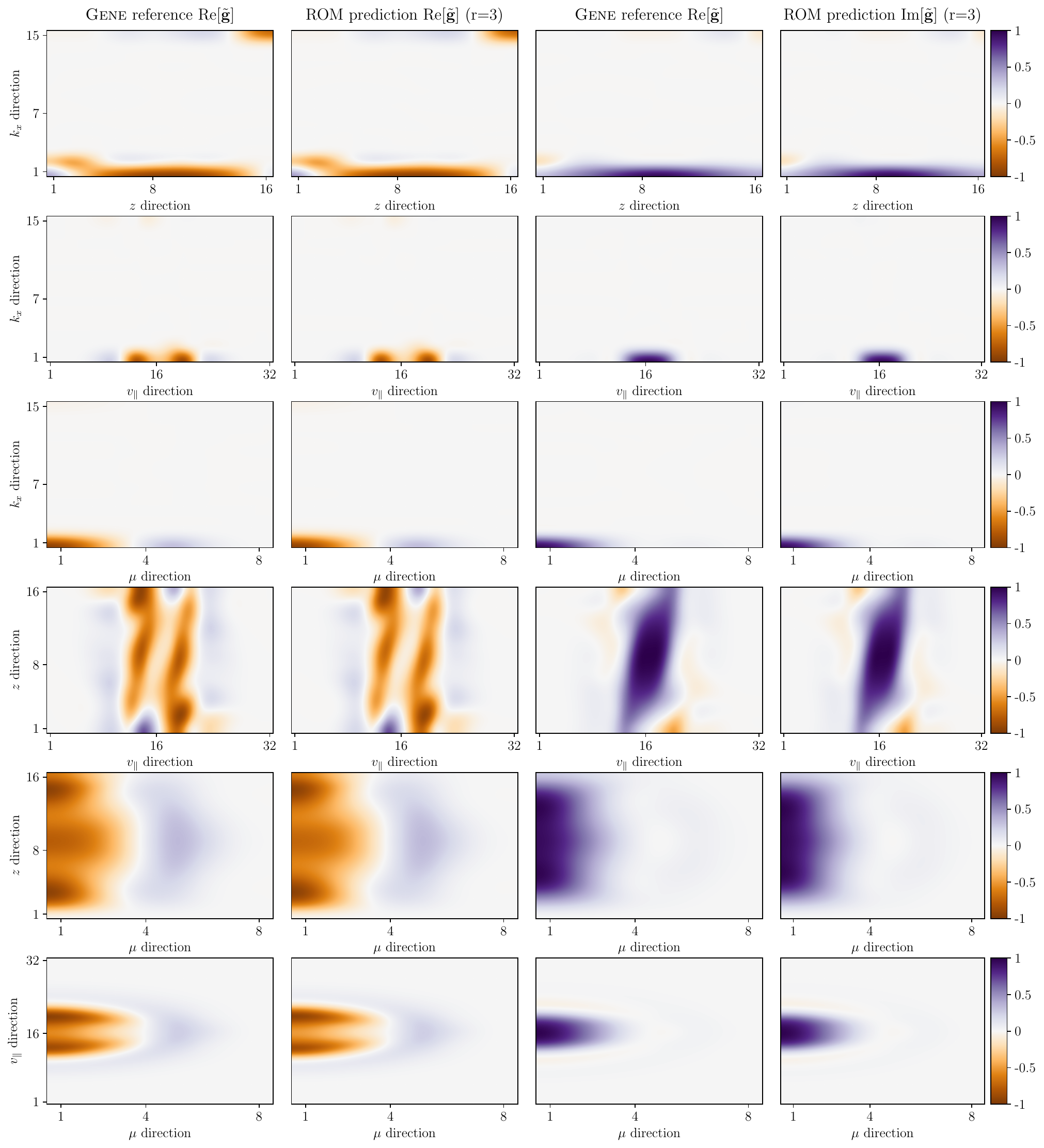}
  \caption{Cyclone Base Case benchmark scenario: Comparison between ROM prediction ($r=3$ and $n_p = 8$ instances) and the \textsc{Gene} reference result for the full time-averaged gyrokinetic distribution function, shown for test case $k_y \rho_s = 0.25$.} 
  \label{fig:CBC_parametric_0025_phase_space}
\end{figure}

\begin{figure}[htbp!]
  \centering 
  \includegraphics[width=1.0\textwidth]{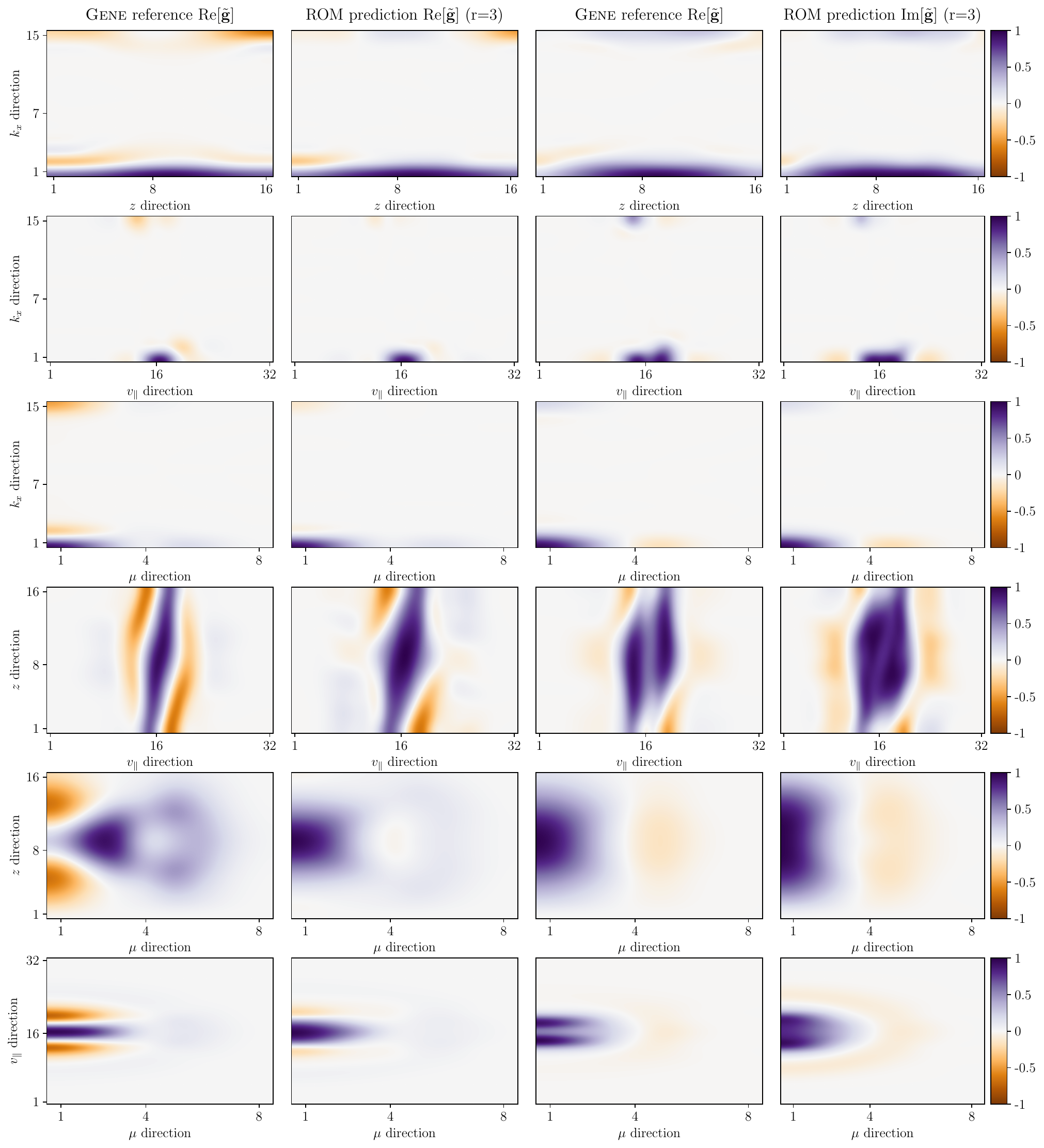}
  \caption{Cyclone Base Case benchmark scenario: Comparison between ROM prediction ($r=3$ and $n_p = 8$ instances) and the \textsc{Gene} reference result for the full time-averaged gyrokinetic distribution function, shown for test case $k_y \rho_s = 0.10$.} 
  \label{fig:CBC_parametric_0010_phase_space}
\end{figure}

Figures~\ref{fig:CBC_parametric_0025_phase_space} and~\ref{fig:CBC_parametric_0010_phase_space} compare all 2D slices of the time-averaged distribution function with $n_{\mathrm{qs}} = 130$.
Figure~\ref{fig:CBC_parametric_0025_phase_space} shows the results for $k_y \rho_s = 0.25$, corresponding to the most accurate growth rate and frequency predictions. 
In contrast, Fig.~\ref{fig:CBC_parametric_0010_phase_space} presents the results for $k_y \rho_s = 0.10$, the testing parameter instance with the largest growth rate and frequency errors.
All 2D slices were generated using bicubic spline interpolation. 
Overall, the ROM captures most of the average features across the different combinations. 
The main discrepancies occur in the $(z, \mu)$ and $(v_{\parallel}, \mu)$ maps for $k_y \rho_s = 0.10$, highlighting pitch-angle anisotropy and late-time parallel/velocity-space structure, respectively. 
By comparison, the $k_y \rho_s = 0.25$ case demonstrates close agreement across all six 2D combinations, illustrating a well-predicted scenario.

To summarize, this section demonstrated that low-dimensional optDMD ROMs can provide fast and reliable predictions both beyond a training time horizon and for variations in the binormal wavenumber $k_y \rho_s$ for the CBC benchmark, characterized by ITG-driven micro-instabilities. 
Next, we assess the capabilities of parametric sparse grid optDMD ROMs in a real-world scenario involving variations in six plasma parameters.

\subsection{Electron temperature gradient-driven micro-instability in the tokamak pedestal}
\label{subsec:res_ETG}
The second scenario is a realistic problem focused on ETG mode simulation in the near-edge pedestal region of the DIII-D experiment, similar to the setup considered in Refs.~\cite{Hassan_2022,walker}.
This is motivated by the study of electron-scale turbulent transport in the near-edge plasma of fusion experiments.
Achieving the high core temperatures and densities required for self-heated (`burning') plasmas in future power plants necessitates the creation and sustainment of a steep gradient region at the plasma edge, known as the pedestal. 
The formation of this pedestal is a complex and still incompletely understood process, where plasma turbulence plays a crucial role. 
Driven by large spatial gradients in density and temperature, turbulence within the pedestal induces transport that contributes to its self-regulation. 
A significant driver of this turbulence is the ETG micro-instability~\cite{Dorland,Jenko_GENE}, which operates on sub-millimeter scales perpendicular to the magnetic field. 
Quantifying the impact of ETG turbulence on the pedestal structure is therefore essential for designing fusion configurations with enhanced energy confinement.

\subsubsection{Setup for high-fidelity gyrokinetic simulations}
We model the plasma behavior using six local parameters: $\{n_e, T_e, \omega_{n_e}, \omega_{T_e}, q, \tau \}$. 
Here, $n_e [10^{19} \mathrm{m^{-3}}]$ is the electron density, and $T_e [\mathrm{keV}]$ is the electron temperature, $\omega _{n_e}=R/L_{n_e}$ and $\omega _{T_e} = R/L_{T_e}$ are the respective normalized (with respect to the major radius $R$) gradients, $q$ is the safety factor, and $\tau=Z_{\rm eff}T_e/T_i$, where $Z_\mathrm{eff}$ is the effective ion charge retained in the collision operator (here, a linearized Landau--Boltzmann operator) and $T_i$ denotes the ion temperature.
The ions are assumed to be adiabatic, and electromagnetic effects, computed consistently with the values of the electron temperature and density, are retained.
Table~\ref{tab:ETG_parameters} summarizes the setup for the six parameters.
These parameters are modeled as uniform random variables with bounds representative of typical experimental error bars.
The first two parameters are varied by $\pm 10\%$ from their nominal values, while the remaining four inputs, including the two gradients, are varied by $\pm 20\%$.
\begin{table}[htbp]
\centering
\begin{tabular}{|c|c|c|c|}
\hline
input parameter & left bound & right bound \\
\hline
electron temperature $T_{e} [\mathrm{keV}]$ & $0.357$ & $0.436$ \\
electron density $n_{e}$ [$10^{19} \mathrm{m^{-3}}$] & $4.042$ & $4.941$ \\
temperature gradient $\omega_{T_e} = R/L_{T_e}$ & $148.800$ & $223.200$ \\
density gradient $\omega_{n_e} = R/L_{n_e}$ & $70.400$ & $105.600$\\
electron-to-ion temperature ratio $\tau$ & $1.152$ & $1.728$ \\
safety factor $q$ & $3.628$ & $5.443$ \\
\hline
\end{tabular}
\caption{Electron temperature gradient-driven micro-instability scenario: Summary of the six input parameters and their respective uniform bounds used to construct parametric reduced models.}
\label{tab:ETG_parameters}
\end{table}

The considered setup models DIII-D pedestal conditions, similar to those in Refs.~\cite{Hassan_2022,walker}.
A comprehensive description of the experimental setup can be found in~\cite{Hassan_2022}.
We note that a similar setup was also considered in Refs.~\cite{FMJ22,FMJ24} where sparse grids were shown to enable efficient uncertainty propagation and sensitivity analysis, as well as the construction of a surrogate transport model for the ETG heat flux directly from nonlinear turbulence simulations.

We specify the magnetic geometry using a generalized Miller parametrization \cite{Candy_2009} using $64$ Fourier harmonics to achieve a sufficiently accurate representation of the flux-surface at $\rho_{tor}=0.95$, near the pedestal top. 
The high-fidelity linear \textsc{Gene} simulations employ $n = n_{k_x}\times n_{k_y}\times n_{z}\times n_{v_\|}\times n_{\mu}=3\times1\times168\times32\times8 = 129,024$ degrees of freedom in the five-dimensional position--velocity space.
Even with acceleration from four NVIDIA A100 GPUs, calculations of the linear growth rate typically require $\sim$$150$s wall clock time.
The main cost arises from requiring a high resolution of points along the magnetic field line to accurately resolve the fast electron dynamics, which is endemic to electron scale instabilities in the pedestal~\cite{Hassan_2022}.

\subsubsection{Setup for sparse-grid-accelerated parametric data-driven reduced modeling}
To learn a parametric optDMD ROM that embeds variations in the six considered plasma parameters, we generate training data using a level $L=3$ sparse grid.
This grid comprises $n_p=28$ $(L)$-Leja points, computed with respect to the uniform distribution whose bounds are given in Table~\ref{tab:ETG_parameters}. 
This approach is substantially more efficient than a full tensor grid: a mere three points per parameter in a tensor grid would require $3^6=729$ \textsc{Gene} simulations, whereas our method uses $26\times$ fewer simulations; cf.~Fig.~\ref{fig:sg_vs_fg}.

The choice of a level $L=3$ sparse grid is motivated as follows.
Since $(L)$-Leja sparse grids are hierarchical, we can start with a grid of level $L=1$ (consisting of a single point) and incrementally increase the level until a target accuracy is achieved.
To evaluate the predictive performance of the resulting sparse-grid-accelerated optDMD ROM, we generate $20$ testing parameter values pseudo-randomly from the same uniform distribution used for training.
The corresponding \textsc{Gene} runtimes are reported in the second column in Table~\ref{tab:etg-runtime-comparison}.  
On four GPUs of a single Perlmutter node, the average \textsc{Gene} runtime was $83.960$ seconds, with a minimum of $54.245$ seconds and a maximum of $161.686$ seconds.

\begin{figure}[htbp!]
  \centering 
  \includegraphics[width=0.7\textwidth]{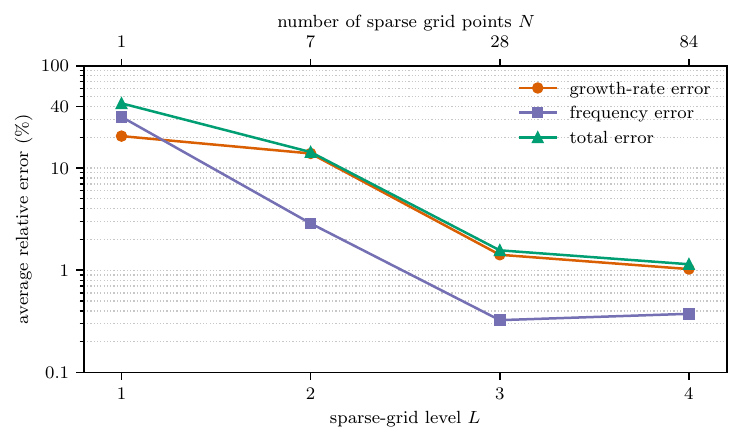}
  \caption{Electron temperature gradient-driven micro-instability scenario: Average relative errors in growth rate, frequency, and combined total error for increasing sparse-grid levels, computed for $20$ out-of-sample testing parameter instances.} 
  \label{fig:sg_error}
\end{figure}

We consider sparse grids at levels $L \in \{1,2,3,4\}$, corresponding to $N = \{1,7,28,84\}$ grid points respectively.
For each resulting ROM, we compute the growth rate and frequency of the dominant ETG mode at all $20$ test points, and evaluate the average relative errors as well as the combined total error.
These errors are plotted in Fig.~\ref{fig:sg_error}.
The results exhibit rapid convergence from $L=2$ to $L=3$, with a $\sim$$9\times$ reduction in total error, followed by diminishing returns from $L=3$ to $L=4$, which yields an additional $\sim$$27\%$ decrease in total error.  
In particular, although the growth rate error decreases by $\sim$$28\%$ from $L=3$ to $L=4$, the frequency error changes only marginally (and slightly increases on average), so the overall benefit of tripling the training cost is modest.  
Based on this trade-off between accuracy and computational expense, we select a level-3 sparse grid (i.e., $n_p=28$ $(L)$-Leja points) as a practical and efficient choice.

\begin{table}[htbp!]
\centering
\begin{tabular}{|c|c|c|c|}
\hline
	 Test Case 
 & \textsc{Gene} runtime (s)
 & ROM runtime (s) 
 & Speedup factor \\
\hline
	  1   & 119.805 & 0.083 & 1,443 \\
    2   & 65.248  & 0.084 & 776 \\
    3   & 71.917  & 0.082 & 877 \\
    4   & 101.922 & 0.087 & 1,171 \\
    5   & 59.434  & 0.080 & 742 \\
    6   & 54.245  & 0.083 & 653 \\
	  7   & 61.259  & 0.082 & 747 \\
    8   & 59.994  & 0.085 & 705 \\
    9   & 161.686 & 0.084 & 1,924 \\
    10  & 107.547 & 0.082 & 1,311 \\
    11  & 78.791  & 0.081 & 972 \\
    12  & 61.078  & 0.083 & 735 \\ 
    13  & 81.540  & 0.081 & 1,006 \\
    14  & 93.739  & 0.084 & 1,115 \\
	  15  & 57.456  & 0.082 & 700 \\
    16  & 100.410 & 0.085 & 1,181  \\
    17  & 65.536  & 0.082 & 799 \\
    18  & 109.453 & 0.083 & 1,318 \\
    19  & 83.044  & 0.081 & 1,025 \\
    20  & 85.091  & 0.082 & 1,037 \\
\hline
Average & 83.960  & 0.083 & 1,011 \\
\hline
\end{tabular}
\caption{Electron Temperature Gradient benchmark scenario: Comparison between the  runtimes of the reference \textsc{Gene} simulations and the optDMD ROM with dimension $r=4$. The ROM offline construction cost is $807$ seconds. The ROM prediction equates to $473$ snapshots over the same time horizon used for training $[t_i, t_f] = [0.0, 0.084]$.}
\label{tab:etg-runtime-comparison}
\end{table}

\subsubsection{Reduced-order model predictions for parameters beyond training}
We begin by selecting the time domains for training the parametric ROM. 
We adopt a common-time policy in which we set $t_i=0$ and define $t_f$ as the shortest available physical duration across the $28$ training runs.
This yields $t_f = 0.084$.
All training snapshots are taken uniformly on the shared interval $t\in[0.000, 0.084]$ with the same step $\Delta t$ (after downsampling by a factor of $10$), yielding an identical count of $n_t=473$ snapshots per parameter instance. 
Out-of-sample forecasts are evaluated on $[0.000, 0.084]$ as well. 
In total this gives $473\times 28=13{,}244$ snapshots and a global snapshot matrix $\mathbf{G}\in\mathbb{C}^{129{,}024\times 13{,}244}$.
This setup is summarized in Table~\ref{tab:etg-training-stats}. 
This table also includes the minimum, maximum, and average \textsc{Gene} runtime, with the average runtime being $91.695$ seconds.
\begin{table}[htbp]
\centering
\begin{tabular}{|c|c|c|c|}
\hline
Setup for acquiring training data & Minimum & Maximum & Average \\ 
\hline
Computed time steps (\textsc{Gene})
  & 4,729 
  & 11,169 
  & 8,174.36 \\

\textsc{Gene} runtime (s)
  & 52.583 
  & 126.431 
  & 91.695 \\
\hline
\multicolumn{4}{|c|}{$[t_i, t_f] = [0.0,\;0.084]$} \\ 
\hline

\hline
\end{tabular}
\caption{Electron temperature gradient-driven micro-instability scenario: Details about the \textsc{Gene} simulations used to acquire the training data.}
\label{tab:etg-training-stats}
\end{table}

\begin{figure}[htbp!]
  \includegraphics[width=1.0\textwidth] {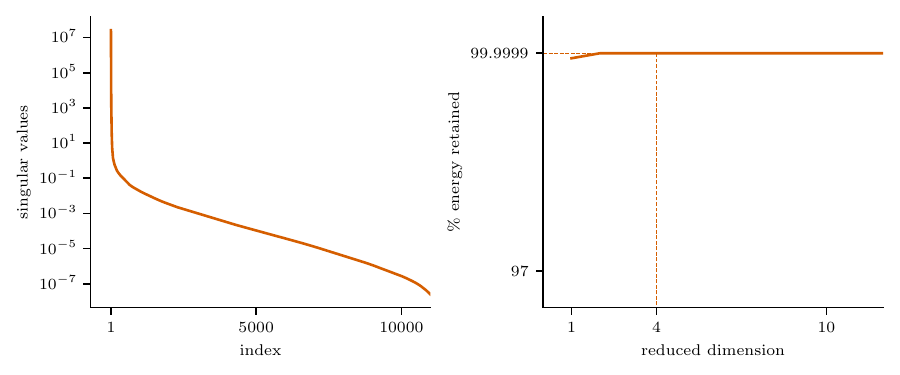}
  \caption{Electron temperature gradient-driven micro-instability scenario: Singular values (left) and corresponding retained energy (right). The training data consists of the \textsc{Gene} distribution function over the respective micro-instability saturation phase for $n_p=28$ sparse $(L)$-Leja grid parameter instances.} 
  \label{fig:ETG-svd}
\end{figure}

We then determine the reduced dimension for the parametric ROM.
Figure~\ref{fig:ETG-svd} plots the singular values and corresponding retained energy, demonstrating that $r=4$ POD modes retain over $99.9999\%$ of the total energy. 
We therefore construct a sparse grid parametric optDMD ROM with a reduced dimension of $r=4$.
The parametric ROM offline cost is $807$ seconds on one Perlmutter CPU node.

Next, we assess the prediction capabilities of this ROM using the $20$ testing parameter values.
The third and fourth columns in Table~\ref{tab:etg-runtime-comparison} report the ROM online evaluation times and speedups, respectively, for all $20$ test cases. 
Our ROM required $0.083$ seconds on average (on a single CPU core) to predict the full distribution function, demonstrating an average speedup factor exceeding $1,000$.

Table~\ref{tab:ETG-lambda-comparison} compares the dominant eigenvalues computed with \textsc{Gene} against those predicted by the ROM for all $20$ out-of-sample test parameter points, reporting the corresponding relative errors.
The vast majority of growth-rate and frequency predictions exhibit relative errors below 
$1\%$.
The largest deviations occur in test cases~9 and~18, where the growth-rate errors are 
$8.715\%$ and $5.229\%$, respectively.
Overall, these results highlight the accuracy of the parametric ROM and demonstrate the effectiveness of sparse-grid interpolation with L-Leja points for capturing parametric variations in ETG modes.

\begin{table}[htbp]
\centering
\begin{tabular}{|c|c|c|c|c|}
\hline
Test Case &  \(\upalpha_{\mathrm{ref}}\) &
$\tilde{{\upalpha}}_{\mathrm{ROM}}$  & Growth Rate Error (\%) & Frequency Error (\%) \\
\hline
1  & \(102.218 - 425.998\)$\mathrm{i}$ & \(100.551-425.623\)$\mathrm{i}$ & 1.630 & 0.088 \\
2  & \(143.815 - 184.055\)$\mathrm{i}$ & \(144.198-186.111\)$\mathrm{i}$ & 0.266 & 1.117 \\
3  & \(151.346 - 227.680\)$\mathrm{i}$ & \(152.400-228.618\)$\mathrm{i}$ & 0.696 & 0.412 \\
4  & \(123.600 - 422.313\)$\mathrm{i}$ & \(125.897-421.017\)$\mathrm{i}$ & 1.858 & 0.307 \\
5  & \(181.105 - 401.320\)$\mathrm{i}$ & \(179.419-401.157\)$\mathrm{i}$ & 0.931 & 0.041 \\
6  & \(179.880 - 322.169\)$\mathrm{i}$ & \(180.497-322.710\)$\mathrm{i}$ & 0.343 & 0.168 \\
7  & \(180.735 - 326.660\)$\mathrm{i}$ & \(181.072-327.666\)$\mathrm{i}$ & 0.187 & 0.308 \\
8  & \(167.890 - 339.058\)$\mathrm{i}$ & \(168.548-338.864\)$\mathrm{i}$ & 0.392 & 0.057 \\
9  & \(67.166  - 598.561\)$\mathrm{i}$ & \(73.020-595.118\)$\mathrm{i}$  & 8.715 & 0.575 \\
10 & \(112.340 - 522.321\)$\mathrm{i}$ & \(114.453-522.102\)$\mathrm{i}$ & 1.881 & 0.042 \\
11 & \(147.970 - 227.259\)$\mathrm{i}$ & \(145.208-227.864\)$\mathrm{i}$ & 1.867 & 0.266 \\
12 & \(163.487 - 224.390\)$\mathrm{i}$ & \(161.522-226.409\)$\mathrm{i}$ & 1.202 & 0.900 \\
13 & \(145.603 - 259.407\)$\mathrm{i}$ & \(144.531-259.162\)$\mathrm{i}$ & 0.736 & 0.095 \\
14 & \(129.197 - 289.963\)$\mathrm{i}$ & \(129.206-289.191\)$\mathrm{i}$ & 0.007 & 0.266 \\
15 & \(180.474 - 300.428\)$\mathrm{i}$ & \(180.149-302.816\)$\mathrm{i}$ & 0.180 & 0.795 \\
16 & \(129.533 - 412.653\)$\mathrm{i}$ & \(130.137-412.337\)$\mathrm{i}$ & 0.467 & 0.076 \\
17 & \(145.947 - 207.040\)$\mathrm{i}$ & \(145.848-208.140\)$\mathrm{i}$ & 0.068 & 0.531 \\
18 & \(103.983 - 590.134\)$\mathrm{i}$ & \(98.473-587.675\)$\mathrm{i}$  & 5.299 & 0.417 \\
19 & \(143.528 - 389.056\)$\mathrm{i}$ & \(145.644-389.009\)$\mathrm{i}$ & 1.474 & 0.012 \\
20 & \(146.416 - 317.500\)$\mathrm{i}$ & \(146.171-317.413\)$\mathrm{i}$ & 0.167 & 0.027 \\
\hline
\end{tabular}
\caption{Electron temperature gradient-driven micro-instability scenario: Comparison between the ROM prediction ($r=4$, trained with $n_p=28$ instances) and the \textsc{Gene} reference result for the eigenvalue corresponding to the dominant ETG micro-instability mode, shown for $20$ parameter instances beyond training.}
\label{tab:ETG-lambda-comparison}
\end{table}

\begin{figure}[htbp!]
  \includegraphics[width=1.0\textwidth]{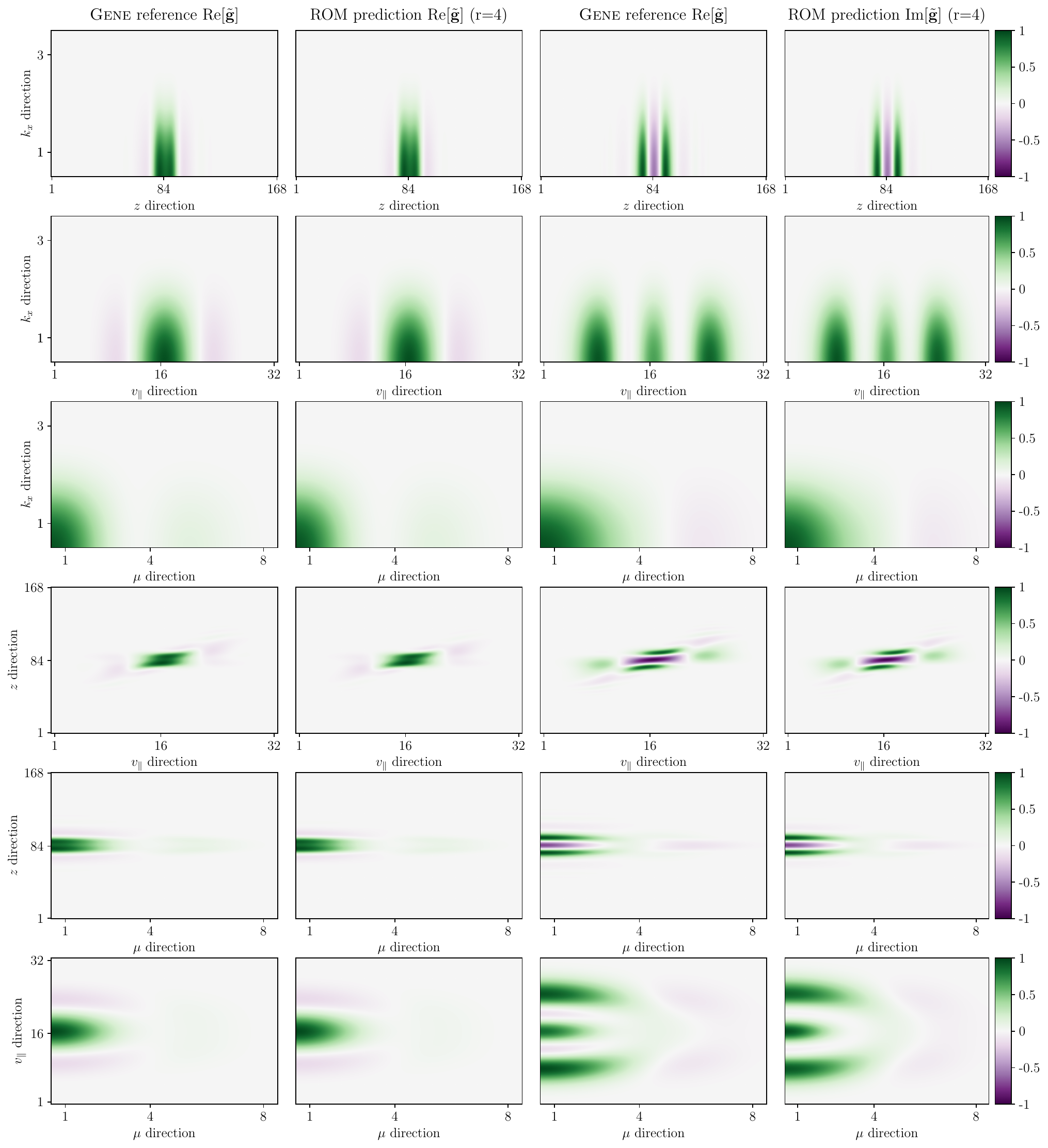}
  \caption{Electron temperature gradient-driven micro-instability scenario: Comparison between ROM prediction ($r=4$) and the \textsc{Gene} reference result for the full time-averaged gyrokinetic distribution function, shown for test case $20$.} 
  \label{fig:ETG_parametric_20_phase_space}
\end{figure}

\begin{figure}[htbp!]
  \includegraphics[width=1.0\textwidth]{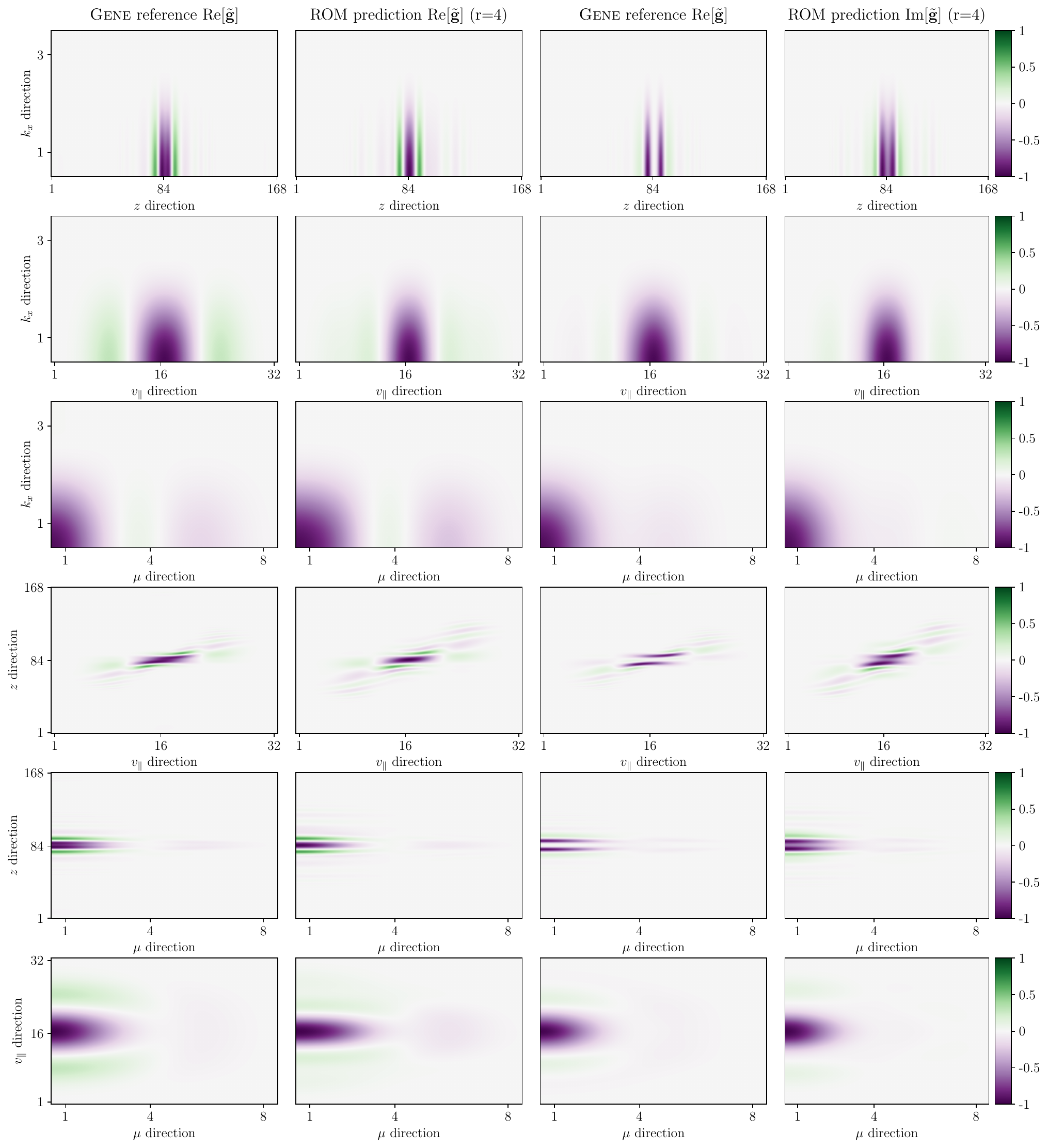}
  \caption{Electron temperature gradient-driven micro-instability scenario: Comparison between ROM prediction ($r=4$) and the \textsc{Gene} reference result for the full time-averaged gyrokinetic distribution function, shown for test case $9$.} 
  \label{fig:ETG_parametric_9_phase_space}
\end{figure}

\begin{figure}[htbp!]
  \centering 
  \includegraphics[width=0.9\textwidth]{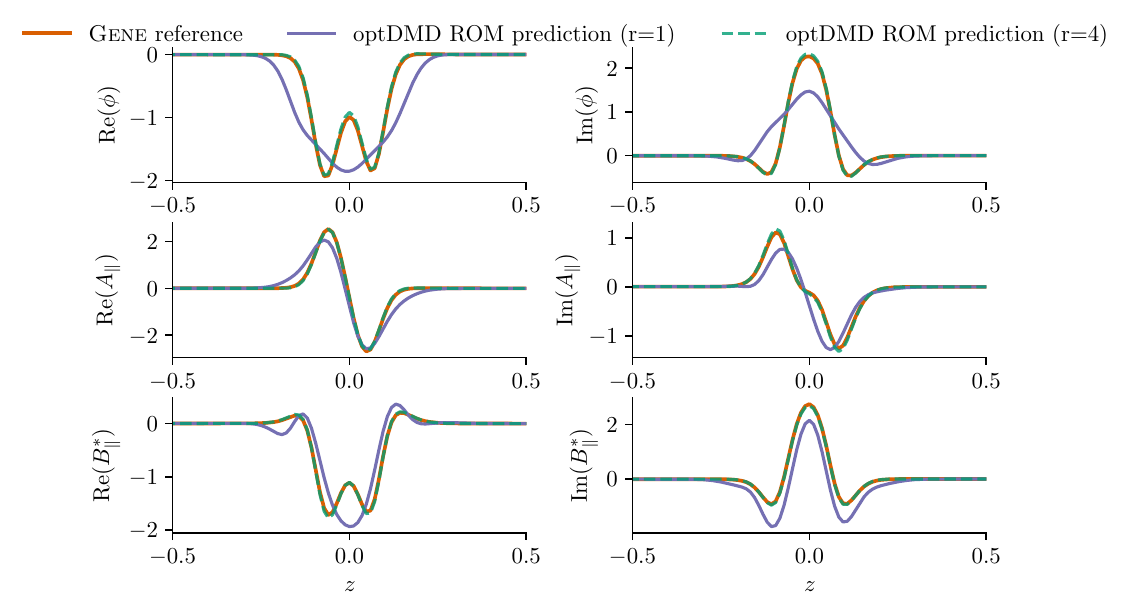}
  \caption{Electron temperature gradient-driven micro-instability scenario: Comparison between the ROM prediction ($r=1$ and $r=4$) and the \textsc{Gene} reference result for the electromagnetic fields (derived from the distribution function) for test case $20$.} 
  \label{fig:ETG_20_fields_multi_r}
\end{figure}

\begin{figure}[htbp!]
  \centering 
  \includegraphics[width=0.9\textwidth]{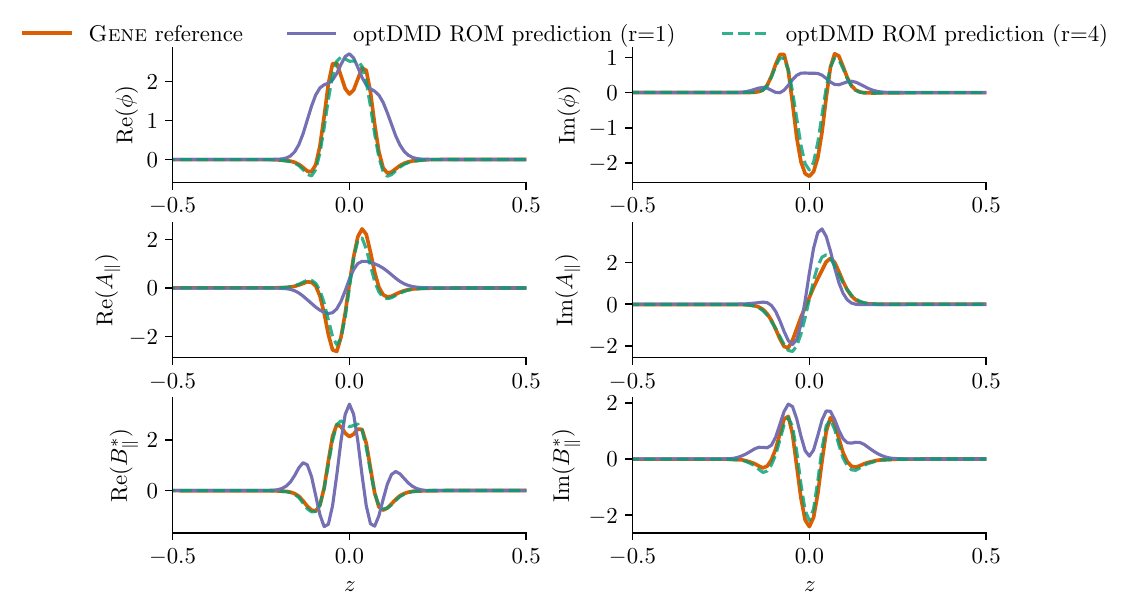}
  \caption{Electron temperature gradient-driven micro-instability scenario: Comparison between the ROM prediction ($r=1$ and $r=4$) and the \textsc{Gene} reference result for the electromagnetic fields (derived from the distribution function) for test case $9$.} 
  \label{fig:ETG_9_fields_multi_r}
\end{figure}

We provide a more detailed assessment of the ROM's prediction capabilities, focusing on testing parameters $20$ and $9$ from Table~\ref{tab:ETG-lambda-comparison}, which correspond to the smallest and largest eigenvalue prediction errors, respectively.
We use the optDMD ROM to predict the full distribution function over $[0.000, 0.084]$.
Figures~\ref{fig:ETG_parametric_20_phase_space} and~\ref{fig:ETG_parametric_9_phase_space} plot all 2D time-averaged slices of the distribution function for cases $9$ and $20$, computed as described in Sec.~\ref{subsubsec:CBC_2D_slices} with $n_{\mathrm{qs}} = 170$. 
For both cases, good agreement is observed between the ROM predictions and the \textsc{Gene} reference results, demonstrating that the ROM captures the averaged features of the distribution function.

We further use the predicted distribution functions to compute other physically relevant quantities.
Specifically, we compute moments of the distribution function, namely the electrostatic potential $\phi$, the parallel component of the vector potential $A_\parallel$, and the parallel component of the effective magnetic field $B^*_\parallel$ along the magnetic field direction $z$, as summarized in Sec.~\ref{subsec:background_gyrokinetic_modeling}.
These quantities are computed over the last $n_{\mathrm{qs}} = 170$ snapshots within $[0.000, 0.084]$.
Given their relevance, we also investigate the impact of the ROM reduced dimension.
To this aim, we consider parametric sparse grid optDMD ROMs with reduced dimensions $r=1$ and $r=4$.
Figures~\ref{fig:ETG_20_fields_multi_r} and~\ref{fig:ETG_9_fields_multi_r} compare the $\phi$, $A_\parallel$, and $B^*_\parallel$ computed using the parametric ROM solution against the reference \textsc{Gene} results for testing parameters $9$ and $20$, respectively. 
For a consistent visualization, the results were normalized such that the inner product of the fields is unity, that is, $\langle \phi, \phi \rangle = 1$.
In both cases, the ROM with reduced dimension $r=1$ fails to capture the electromagnetic fields, whereas the ROM with reduced dimension $r=4$ produces much more reliable predictions.
This demonstrates that, although the first POD mode contains a large fraction of the system’s energy, a larger number of modes may be required to construct ROMs that produce physically consistent, predictive results.

\begin{figure}[htbp!]
  \centering 
  \includegraphics[width=0.7\textwidth]{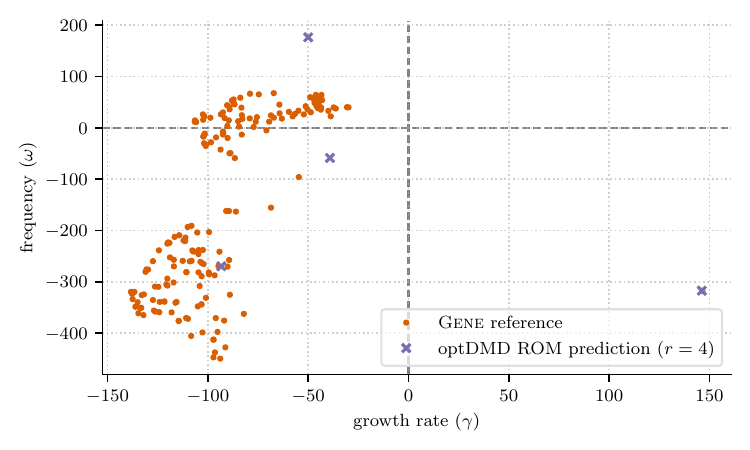}
  \caption{Electron temperature gradient-driven micro-instability scenario: Comparison between the ROM prediction ($r=4$) and the \textsc{Gene} reference result for the eigenvalue spectrum for test case $20$. } 
  \label{fig:ETG_20_eigs}
\end{figure}

\begin{figure}[htbp!]
  \centering 
  \includegraphics[width=0.7\textwidth]{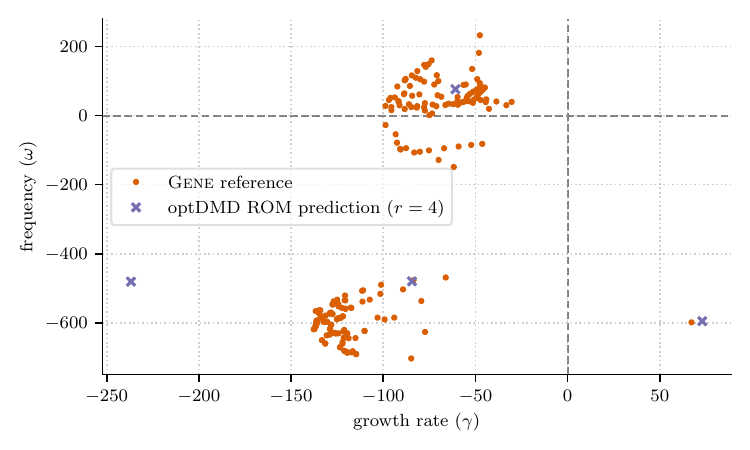}
  \caption{Electron temperature gradient-driven micro-instability scenario: Comparison between the ROM prediction ($r=4$) and the \textsc{Gene} reference result for the eigenvalue spectrum for test case $9$.} 
  \label{fig:ETG_9_eigs}
\end{figure}

Finally, for the computed parametric ROM with reduced dimension $r=4$, we investigate its $r-1=3$ predicted stable eigenmodes. 
In general, practitioners are interested in these stable modes for their ability to affect saturation and transport via nonlinear coupling mechanisms~\cite{terry2006role}. 
Therefore, being able to predict not only the dominant instability but also the subdominant and stable eigenmodes is a desirable property of parametric ROMs. 
Figures~\ref{fig:ETG_20_eigs} and~\ref{fig:ETG_9_eigs} plot the the four eigenvalues predicted by the parametric ROM compared with the $\sim$$200$ eigenvalues as computed by the \textsc{Gene} linear eigenvalue solver (details discussed in Sec.~\ref{subsec:plasma_setup}), for testing parameters $20$ and $9$. 
For both cases, the growth rate and frequency of the most unstable mode is recovered with high accuracy.  
Since the \textsc{Gene} linear eigenvalue solver produces many more stable modes than our ROM, a direct one-to-one comparison of all stable modes is nontrivial. 
To make the comparison meaningful, we restrict attention to targeted regions of the stable spectrum using shift–and–invert preconditioning and then match ROM modes to \textsc{Gene} modes within that window using nearest-neighbor distance in the complex plane.
Taking test case 9 as an example, one can see that the stable mode predicted by our optDMD ROM $\tilde{{\upalpha}}_{\mathrm{ROM}} = -84.442 -479.979 \text{i}$ can be matched (with reasonable confidence) to the stable mode computed by the eigenvalue solver, $\upalpha_{\mathrm{ref}} = -83.604-476.662\text{i}$, with $1.002\%$ (growth) and $0.696\%$ (frequency) relative errors. 
For other stable eigenvalues (e.g., $\tilde{\alpha}_{\mathrm{ROM}}=-60.871+76.517\,\mathrm{i}$ versus $\alpha_{\mathrm{ref}}=-49.599+74.413\,\mathrm{i}$), the mapping is more ambiguous, yielding $22.726\%$ and $2.827\%$ relative errors in growth and frequency, respectively.
Possible explanations for these observed discrepancies in the stable modes include: (i) in the damped region, nearest-neighbor assignments become fragile; (ii) dimensionality reduction with $r=4$ compress many stable directions into a low-dimensional subspace, so the predicted optDMD spectrum need not coincide exactly with the eigen-spectrum of the linear operator from \textsc{Gene}; and (iii) fitting on a particular (common) time horizon directly impacts which stable modes are recovered. 
Thus, although a direct identification of all stable modes may be hard to conclude, visual inspection suggests that there are many possible candidate modes that may correspond to the stable modes recovered by the ROM. 
Furthermore, in contrast to cases studied elsewhere~\cite{Faber2018, Hatch2013, Pueschel2016, Whelan2018}, for the ETG case currently considered, these subdominant and stable modes evidently contribute with very little energy to the overall spectrum; therefore, we leave further exploration of stable and subdominant modes for future work, while noting that the proposed approach remains promising.

\section{Conclusion} \label{sec:conclusion}
Parametric data-driven reduced-order modeling plays a key role in enabling tasks such as design optimization, control, uncertainty quantification, sensitivity analysis, and inverse problems, particularly for applications involving large-scale numerical models.  
A central challenge is to ensure that the training cost of these reduced models remains manageable.  
One effective approach is to employ efficient techniques for selecting training parameter instances.  
This paper investigated the use of sparse grids constructed with $(L)$-Leja points for efficiently training parametric reduced models.  
Since this point sequence is nested and grows slowly with the sparse grid level, the cardinality of the resulting sparse grid remains reasonably low even in higher-dimensional settings. 
Moreover, $(L)$-Leja points are well-suited for interpolation, enabling the computation of predictions for arbitrary input parameter values within the interpolation domain.  
We demonstrate the effectiveness of this approach in plasma micro-instability simulations.  
For a real-world scenario involving six key plasma parameters associated with electron temperature gradient driven modes at the edge of fusion experiments, only 28 parameter instances were sufficient to construct an effective parametric reduced model based on optimized dynamic mode decomposition.  
The ideas presented here are generic and extend beyond plasma micro-instability simulations and linear parametric reduced models.  
A potential limitation arises in applications with computationally very-expensive models: the sparse $(L)$-Leja grid may contain a large number of points if the input parameter space is high-dimensional.  
This can be mitigated by incorporating adaptivity into the sparse grid procedure, which has been shown to be effective for uncertainty propagation~\cite{FMJ22, FDJ21} and surrogate model construction~\cite{FMJ24}.
Moreover, the proposed parametric reduced-order modeling workflow is intended for predictions in the interpolative regime only.
Extending it to support reliable extrapolation is left for future research.

\section{Acknowledgment} \label{sec:acknowledgement}
This research used resources of the National Energy Research Scientific Computing Center (NERSC), a U.S. Department of Energy Office of Science User Facility located at Lawrence Berkeley National Laboratory, operated under Contract No.~DE-AC02-05CH11231 using NERSC award FES-ERCAP0026282.
I.-G.~F. was supported in part by NSF Grant DMS-2436357 ``MATH-DT: Advancing Digital Twins for Jet Engines Through Mathematical and Computational Innovation." 
B.~J.~F. was supported in part by U.S. Department of Energy Grant No. DE-FG02-93ER54222. 

\begin{appendix}
\section{Parameter setup for the electron temperature gradient-driven scenario}
\label{sec:etg-parameters}

Table~\ref{tab:ETG-train-params} summarizes the $n_p=28$ training parameter instances generated using a level-3 sparse grid with $(L)$-Leja points.
Table~\ref{tab:ETG-test-params} presents the $20$ testing parameters generated pseudo-randomly from the same uniform distribution used for training (cf.~Table~\ref{tab:ETG_parameters}).
\begin{table}[h]
    \centering
    \begin{tabular}{|c|c|c|c|c|c|c|}
        \hline
        Training \# &
        \makecell[c]{electron\\temperature \\ $T_{e}\ [\mathrm{keV}]$} &
        \makecell[c]{electron\\density $n_{e}$ \\ $[10^{19}\ \mathrm{m^{-3}}]$} &
        \makecell[c]{temperature\\gradient \\ $\omega_{T_e} = R/L_{T_e}$} &
        \makecell[c]{density\\gradient \\ $\omega_{n_e} = R/L_{n_e}$} &
        \makecell[c]{electron-to-ion\\temperature ratio \\ $\tau$} &
        \makecell[c]{safety\\factor \\ $q$} \\
        \hline
1 & 0.397 & 4.492 & 186.000 & 88.000 & 1.440 & 4.536 \\
2 & 0.397 & 4.492 & 186.000 & 88.000 & 1.728 & 4.536 \\
3 & 0.397 & 4.492 & 186.000 & 88.000 & 1.440 & 5.443 \\
4 & 0.397 & 4.492 & 186.000 & 105.600 & 1.440 & 4.536 \\
5 & 0.397 & 4.492 & 223.200 & 88.000 & 1.440 & 4.536 \\
6 & 0.436 & 4.492 & 186.000 & 88.000 & 1.440 & 4.536 \\
7 & 0.397 & 4.941 & 186.000 & 88.000 & 1.440 & 4.536 \\
8 & 0.397 & 4.492 & 186.000 & 88.000 & 1.152 & 4.536 \\
9 & 0.397 & 4.492 & 186.000 & 88.000 & 1.728 & 5.443 \\
10 & 0.397 & 4.492 & 186.000 & 105.600 & 1.728 & 4.536 \\
11 & 0.397 & 4.492 & 223.200 & 88.000 & 1.728 & 4.536 \\
12 & 0.436 & 4.492 & 186.000 & 88.000 & 1.728 & 4.536 \\
13 & 0.397 & 4.941 & 186.000 & 88.000 & 1.728 & 4.536 \\
14 & 0.397 & 4.492 & 186.000 & 88.000 & 1.440 & 3.628 \\
15 & 0.397 & 4.492 & 186.000 & 105.600 & 1.440 & 5.443 \\
16 & 0.397 & 4.492 & 223.200 & 88.000 & 1.440 & 5.443 \\
17 & 0.436 & 4.492 & 186.000 & 88.000 & 1.440 & 5.443 \\
18 & 0.397 & 4.941 & 186.000 & 88.000 & 1.440 & 5.443 \\
19 & 0.397 & 4.492 & 186.000 & 70.400 & 1.440 & 4.536 \\
20 & 0.397 & 4.492 & 223.200 & 105.600 & 1.440 & 4.536 \\
21 & 0.436 & 4.492 & 186.000 & 105.600 & 1.440 & 4.536 \\
22 & 0.397 & 4.941 & 186.000 & 105.600 & 1.440 & 4.536 \\
23 & 0.397 & 4.492 & 148.800 & 88.000 & 1.440 & 4.536 \\
24 & 0.436 & 4.492 & 223.200 & 88.000 & 1.440 & 4.536 \\
25 & 0.397 & 4.941 & 223.200 & 88.000 & 1.440 & 4.536 \\
26 & 0.357 & 4.492 & 186.000 & 88.000 & 1.440 & 4.536 \\
27 & 0.436 & 4.941 & 186.000 & 88.000 & 1.440 & 4.536 \\
28 & 0.397 & 4.043 & 186.000 & 88.000 & 1.440 & 4.536 \\           
    \hline
    \end{tabular}
    \caption{Electron temperature gradient-driven micro-instability scenario: parameter values used for training.}
    \label{tab:ETG-train-params}
\end{table}

\begin{table}[h]
    \centering
    \begin{tabular}{|c|c|c|c|c|c|c|}
        \hline
        Test \# &
        \makecell[c]{electron\\temperature \\ $T_{e}\ [\mathrm{keV}]$} &
        \makecell[c]{electron\\density $n_{e}$ \\ $[10^{19}\ \mathrm{m^{-3}}]$} &
        \makecell[c]{temperature\\gradient \\ $\omega_{T_e} = R/L_{T_e}$} &
        \makecell[c]{density\\gradient \\ $\omega_{n_e} = R/L_{n_e}$} &
        \makecell[c]{electron-to-ion\\temperature ratio \\ $\tau$} &
        \makecell[c]{safety\\factor \\ $q$} \\
        \hline
1 & 0.436 & 4.934 & 200.292 & 95.571 & 1.344 & 5.437 \\
2 & 0.429 & 4.494 & 157.512 & 96.990 & 1.644 & 3.755 \\
3 & 0.434 & 4.310 & 173.414 & 101.703 & 1.427 & 4.057 \\
4 & 0.398 & 4.166 & 218.252 & 96.555 & 1.652 & 5.149 \\
5 & 0.379 & 4.565 & 214.949 & 99.303 & 1.537 & 3.765 \\
6 & 0.415 & 4.616 & 177.669 & 88.744 & 1.385 & 3.630 \\
7 & 0.412 & 4.476 & 179.472 & 90.972 & 1.152 & 3.956 \\
8 & 0.408 & 4.334 & 163.901 & 72.638 & 1.636 & 3.845 \\
9 & 0.422 & 4.827 & 217.968 & 81.354 & 1.213 & 5.312 \\
10 & 0.403 & 4.550 & 205.447 & 76.964 & 1.370 & 4.767 \\
11 & 0.382 & 4.430 & 156.307 & 86.960 & 1.636 & 4.469 \\
12 & 0.388 & 4.271 & 150.078 & 81.752 & 1.719 & 3.920 \\
13 & 0.428 & 4.119 & 167.387 & 89.976 & 1.506 & 4.489 \\
14 & 0.425 & 4.715 & 156.036 & 79.256 & 1.480 & 4.735 \\
15 & 0.411 & 4.704 & 186.413 & 99.674 & 1.316 & 3.644 \\
16 & 0.395 & 4.194 & 208.466 & 94.208 & 1.452 & 5.278 \\
17 & 0.427 & 4.513 & 151.305 & 87.882 & 1.634 & 3.884 \\
18 & 0.415 & 4.490 & 218.815 & 71.690 & 1.595 & 4.232 \\
19 & 0.393 & 4.737 & 218.710 & 104.546 & 1.693 & 4.272 \\
20 & 0.369 & 4.595 & 173.165 & 86.466 & 1.390 & 5.064 \\  
    \hline
    \end{tabular}
    \caption{Electron temperature gradient-driven micro-instability scenario: parameter values used for testing.}
    \label{tab:ETG-test-params}
\end{table}

\end{appendix}

\bibliographystyle{elsarticle-harv} 
\bibliography{parametric_SG_ROMs}
\end{document}